\documentclass[review]{elsarticle}
 \biboptions{comma,sort&compress}

\usepackage[table]{xcolor}

\usepackage{graphicx}
\usepackage{amsmath}
\usepackage{here}
\usepackage{cuted}

\usepackage{here}
\usepackage[dvips]{epsfig}
\definecolor{bluette}{rgb}{.2,.4,0}
\definecolor{salmon}{rgb}{.9,0.68,0.5}
\definecolor{motive}{rgb}{0.2,1,.5}
\definecolor{list}{rgb}{0.3,.8,.1}
\definecolor{moe}{rgb}{1,.7,.5}
\definecolor{mote}{rgb}{.7,.5,.6}
\definecolor{pisello}{rgb}{.1,1,0}
\definecolor{orange}{rgb}{1,.7,0}
\definecolor{oliva}{rgb}{.1,.5,0.3}
\definecolor{greenda}{rgb}{0,.3,.2}
\definecolor{greenli}{rgb}{0.5,.8,.0}
\definecolor{blueda}{rgb}{0,.1,.6}
\definecolor{purple}{rgb}{.7,.1,.2}
\definecolor{marrone}{rgb}{1,0.7,0}
\definecolor{pinky}{rgb}{1,0.8,0.8}
\definecolor{rose}{rgb}{1,0.4,0}

\def\beq{\begin{equation}}
\def\eeq{\end{equation}}
\def\bea{\begin{eqnarray}}
\def\eea{\end{eqnarray}}
\def\bq{\begin{quote}}
\def\eq{\end{quote}}

\def\nnb{\nonumber}
\def\ga{\left(}
\def\dr{\right)}

\def\lrar{\Longrightarrow}
																
\def\nnb{\nonumber}

\def\la{\langle}
\def\ra{\rangle}

\def\ba{\vspace*{-0.2cm}\begin{array}}
\def\ea{\end{array}\vspace*{-0.2cm}}

\def\b{$\bullet~$}
\def\d{$\diamond~$}

\def\als{\alpha_s}

\def\gg2{\la\alpha_s G^2 \ra}
\def\gg3{g^3f_{abc}\la G^aG^bG^c \ra}
\def\ggg4{\la\als^2G^4\ra}

\def\gg{\lag g^{2}_{s} G^2 \rag}
\def\ggg{\lag g^{3}_{s}G^3\rag}

\usepackage{amsmath}
\usepackage{slashed}
\usepackage{color}

\begin{document}
\begin{frontmatter}

\title{QCD parameters and SM-high precisions from $e^+e^-\to$  Hadrons}
\author{Stephan Narison
}
\address{Laboratoire
Univers et Particules de Montpellier (LUPM), CNRS-IN2P3, \\
Case 070, Place Eug\`ene
Bataillon, 34095 - Montpellier, France\\
and\\
Institute of High-Energy Physics of Madagascar (iHEPMAD)\\
University of Ankatso, Antananarivo 101, Madagascar}
\ead{snarison@yahoo.fr}


\date{\today}
\begin{abstract}
\noindent
Using the PDG 22 compilation of the $e^+e^-\to$  Hadrons $\oplus$ the recent CMD3 data for the pion form factor and the value of gluon condensate $\la\alpha_s G^2\ra$ from heavy quarkonia, we extract the value of the four-quark condensate\,:  $\rho\alpha_s\la\bar\psi\psi\ra^2= (5.98\pm 0.64)\times 10^{-4}$ GeV$^6$  and 
the dimension eight condensate\,: $d_8= (4.3\pm 3.0)\times 10^{-2}$ GeV$^8$
from the ratio ${\cal R}_{10}$ of Laplace sum rules to order $\alpha_s^4$.  We show the inconsistency in using at the same time the standard SVZ value of the gluon and the vacuum saturation of the four-quark condensates.  Using the previous values of the four-quark and $d_8$ condensates, we re-extract $\la\alpha_s G^2\ra$ from ${\cal R}_{10}$ to be\,:  $(6.12\pm 0.61)\times 10^{-2}$ GeV$^4$ in perfect agreement with the one from heavy quarkonia. We also use
the lowest $\tau$-like decay moment ${\cal R}_\tau^{ee}$ to extract the value of the QCD coupling. We obtain to  ${\cal O}(\alpha_s^4)$ the mean of FO (fixed order) and CI (contour improved)  PT series within the standard OPE : $\alpha_s(M_\tau)=0.3385(50)(136)_{syst}$ where the last error is an added conservative systematic from the distance of the mean to  the FO and CI central values. Assuming that the $\alpha_s$-coefficients of the PT series grow geometrically as observed in the calculated case, we obtain to ${\cal O}(\alpha_s^5)$ :   $\alpha_s(M_\tau)=0.3262(37)(78)_{syst}$.  These  results lead to\,: $\alpha_s(M_Z)$=0.1207(17)(3) [resp. 0.1193(11)(3)] to order $\alpha_s^4$ [resp. $\alpha_s^5]$ . The corresponding value of the sum of the non-perturbative contribution at $M_\tau$ is\,: $\delta^V_{NP}(M_\tau)=(2.3\pm 0.2)\times 10^{-2}$. Reciprocally, using $\alpha_s(M_\tau)$, $\la\alpha_s G^2\ra$ and $d_8$ as inputs, we test the stability of the value of the four-quark condensate obtained from the lowest $\tau$-like moment. 
We complete our analysis by updating our previous determinations of the lowest order hadronic vacuum polarization  contributions to the lepton anomalies and to $\alpha(M^2_Z)$.  We obtain in Table\,\ref{tab:amu1}\,: $a_\mu\vert^{hvp}_{l.o}= (7036.5\pm 38.9)\times10^{-11}, \, a_\tau\vert^{hvp}_{l.o}= (3494.8\pm 24.7)\times10^{-9} $ and $\alpha^{(5)}(M_Z)\vert_{had}=(2766.3\pm 4.5)\times 10^{-5}$. This new value of $a_\mu$ leads to : $\Delta a_\mu\equiv a_\mu^{exp}-a_\mu^{th} = (142\pm 42_{th}\pm 41_{exp})\times 10^{-11}$ which reduces the tension between the SM prediction and experiment.
\end{abstract}
\end{frontmatter}
\newpage
\section{Introduction}
\vspace*{-0.2cm}
Precise determinations of the QCD parameters are important inputs for testing the Standard Model (SM). Moreover, 
the study of hadronic properties necessitates a control of the non-perturbative effects where the exact solution is analytically lacking due to our poor knowledge of confinement. Some alternative approaches have been proposed such the numerical lattice simulations. Another approximate approach proposed by Shifman-Vainshtein and Zakharov (SVZ)\,\,\cite{SVZa}\,\footnote{For reviews, see e.g.\,\cite{ZAKA,SNB1,SNB2,SNREV1,SNREV2,SNB3,RRY,DERAF,BERTa,YNDB,PASC,DOSCH,COL}.}$^,$\footnote{The Laplace transform property of the SVZ Borel sum rule has been noticed by\,\cite{SNR} while its non-relativistic form has been studied in\,\cite{BELLa}. }  is to parametrize the non-perturbative effects by a sum of vacuum condensates with increasing dimensions within the Operator Product Expansion (OPE). 

In this paper, we shall extract some of these non-perturbative condensates and the QCD coupling from the $e^+e^-\to$ hadrons data using the SVZ Laplace transform and the lowest $\tau$-decay moment sum rules.  We complete the analysis for updating our previous determinations\,\cite{CALMET,SN76,SN78,SNamu,SNalfa} of the lowest order hadronic vacuum polarization  contributions to the lepton anomaly  $a_{\mu,\tau}$ and to the QED coupling $\alpha(M_Z^2)$.

\section{The SVZ - anatomy of the two-point vector correlator}
\vspace*{-0.2cm}

 In this paper, we take  the example of the hadronic two-point correlator built from the $I=1$  vector current :
\beq
 J^\mu_H(x)=\frac{1}{2}{[}: \bar\psi_u\gamma^\mu\psi_u-\bar\psi_d\gamma^\mu\psi_d:{]}.
 \eeq
The two-point function reads: 
\beq
\hspace*{-0.6cm} \Pi^{\mu\nu}_H(q^2\equiv-Q^2)=i\hspace*{-0.1cm}\int \hspace*{-0.15cm}d^4x ~e^{-iqx}\la 0\vert {\cal T} {J^\mu_H}(x)\ga {J^\nu_H}(0)\dr^\dagger \vert 0\ra=-(g^{\mu\nu}q^2-q^\mu q^\nu)\Pi_H(Q^2)
 \label{eq:2-point}
 \eeq
The corresponding SVZ-expansion reads:
\beq
8\pi^2\Pi_H(-Q^2,m_q^2,\mu)=\sum_{D=0,2,4,..}\hspace*{-0.25cm}\frac{C_{D}(Q^2,m_q^2,\mu)\la O_{D}(\mu)\ra}{(Q^2)^{D/2}}~, 
\label{eq:ope}
\eeq
where $\mu$ is the subtraction scale which separates the long and short distance dynamics and $m_q$ is the quark mass. In addition to the usual perturbative contribution, there are QCD vacuum condensates $\la O_{D}(\mu)\ra$\ effects which are assumed by SVZ to approximate  the not yet under good control QCD confinement.  $C_{D}(\mu)$ are separated calculable Wilson coefficients in perturbative theory (PT).  The renormalization of the $D=4$ condensates has been studied in\,\cite{TARRACH,ESPRIU}, of $D=6$ in\,\cite{SNTARRACH} and of $D=8$ in\,\cite{BAGAN}.  In this paper we assume that the observables which we shall look for are described with a good approximation by the condensates up to dimension $D=8$ and neglect the contribution of higher dimensions one. In this way, the $D=8$ condensates which we shall retain is an effective condensate which we expect to absorb into it the sum of the contributions of all higher dimension ones. Then, we shall be concerned with the following quantities:
\subsection*{\b D=0 perturbation theory}
The expression of the two-point function is known to order $\alpha_s^3$\,\cite{LARIN} and to order $\alpha_s^4$\cite{CHET4}. It reads for 3 flavours:
\beq
8{\pi}\,{\rm Im}\Pi_H(t)=  1+a_s+1.6398a_s^2-10.2839a_s^3-106.8798a_s^4+{\cal O}(a_s^5),
\label{eq:pt}
\eeq
where : 
\beq
a_s\equiv \frac{\alpha_s}{\pi}=\frac{2}{-\beta_1{\rm Log} (t/\Lambda^2)}+\cdots,
\eeq 
where $-\beta_1=(1/2)(11-2n_f/3)$ is the first coefficient of the $\beta$-function and $n_f$ is the number of quark flavours;  $\cdots$ stands for higher order terms which can e.g. be found in \cite{SNB2}. We shall use the expression of $\alpha_s$ up to order $\alpha_s^3$ and the value $\Lambda =(342\pm 8)$ MeV  for  $n_f=3$\:\, from the PDG average\,\cite{PDG}.
\subsection*{\b D=2  operators}
Within the standard OPE, the $D=2$ operators concern the quark mass squared which reads:
\beq
d_2\vert{m_q}\equiv C_2\la O_2\ra\vert_{m_q} = -3(\bar m_u^2+\bar m_d^2)\ga 1+\frac{2}{3}a_s\dr.
\eeq
However, it was emphasized in Res.\,\cite{CNZa,SZ,CNZb,D2rev} that some unsual term due to tachyonic gluon mass related to the linear term of the QCD potential may appear in the OPE as a phenomenological parametrization of the non-calculated higher order PT terms and can be an alternative to the large $\beta$-UV renormalon approach. Its  also appears in some ADS-QCD models\,\cite{ADS1,ADS3}. It reads\,\cite{CNZa}:
\beq
d_2\vert_{tach}\equiv C_2\la O_2\ra\vert_{tach} = \ga \frac{32}{2}-8\zeta_3\dr \alpha_s\lambda^2\,\log\frac{Q^2}{\mu^2}. 
\eeq
The value of the tachyonic mass extracted from the $e^+e^-\to$ I=1 hadrons data  and $\pi$ sum rules is\,\cite{SN93,SN95,SNTAU,TERAYEV}:
\beq
\alpha_s\lambda^2\simeq -(7\pm 3)\times 10^{-2}~{\rm GeV}^2.
\eeq
However, to avoid some misinterpretation of the result obtained in this paper, we shall not consider this term in our analysis
but only show its effect in the sum rule. 
\subsection*{\b D=4  light quark condensate}
Its value comes from the well-known Gell-Mann-Oakes Renner pion PCAC relation:
\beq
 (m_u+m_d)\la\bar \psi_u\psi_u+\bar \psi_d\psi_d\ra =-m_\pi^2f_\pi^2\,\, :\,\,\,\,\,\,\,\,\, f_\pi=92.2~{\rm MeV}.
\eeq
Its contribution to the two-point function reads:
\beq
d_4\vert_{\la\bar\psi\psi\ra}\equiv C_4\la O_4\ra\vert_{\la\bar\psi\psi\ra}=4\pi^2\ga 1+\frac{a_s}{3}\dr\la m_u\bar \psi_u\psi_u+m_d\bar \psi_d\psi_d\ra
\eeq
\subsection*{\b  D=4 gluon condensate}
 It plays an important role in the heavy quark sector as the heavy quark vacuum condensate $\la\bar QQ\ra$ vanishes 
 as the inverse of its mass $1/M_Q$. Its  recent value from heavy quarkonia is\,\cite{SNparam}\,\footnote{Different determinations from light and heavy quark channels are reviewed in this paper.}\,:
\beq
\la\alpha_s G^2\ra= (6.49\pm 0.35)\times 10^{-2}~{\rm GeV}^4,
\label{eq:asg2}
\eeq
which is 1.7 times the original SVZ estimate. This result consolidates the previous claims that its size has been underestimated\,\cite{BELLa,BERTa,FESR,LNT}. 
Its contribution  is:
\beq
d_4\vert_{\la G^2\ra}\equiv C_4\la O_4\ra\vert_{\la G^2\ra}=\frac{\pi}{3}\la\alpha_s G^2\ra \ga 1+\frac{7}{6}a_s\dr 
\eeq
\subsection*{\b  D=5 mixed quark-gluon condensate and $D=6$ three-gluon condensate}
Notice that the $D=5$ mixed quark-gluon: $\la\bar\psi\sigma^{\mu\nu}\frac{\lambda_a}{2}G_{\mu\nu}^a\psi\ra$ does not contribute in the chiral limit $m_q=0$ while the
$D=6$  three-gluon:   $ \la g^3f_{abc} G^a_{\mu\rho}G^{b,\rho}_{\nu}G^{c,\nu}_\rho\ra$ do not contribute to leading order in $\alpha_s$\,\cite{MALLIK} such that we shall neglect them.

\subsection*{\b  D=6 four-quark condensate }
It has the generic form :$\la\bar\psi_i\Gamma_1\psi_j\bar\psi_k\Gamma_1\psi_l\ra$ where using a  vacuum saturation 
assumption can be written in a simple way as $\rho\la\bar\psi\psi\ra^2 $ where $\rho$ indicates the deviation from the vacuum saturation assumption. Present analysis indicates that 
\beq
\rho\alpha_s\la\bar\psi\psi\ra^2 = 5.8(9)\times 10^{-4}\,{\rm GeV}^4,
\label{eq:psi2}
\eeq
indicating a huge violation of the vacuum saturation or factorization:
\beq
\alpha_s\la\bar\psi\psi\ra^2\vert_{fac} = 1.0(9)\times 10^{-4}\,{\rm GeV}^4,
\label{eq:fac}
\eeq
by a factor about 5.8. Notice that the four-quark condensate has a mild dependence of $Q^2$ due to the almost cancellation of the $\log(Q^2/\Lambda^2)$ of the running quark condensate and of the one of $\alpha_s$. Its contribution is:
\beq
d_6\equiv C_6\la O_6\ra\vert_{\la\bar\psi\psi^2\ra}= -\frac{896}{81}\pi^3\rho\alpha_s\la\bar \psi\psi\ra^2.
\label{eq:d6}
\eeq
\subsection*{\b  D=8  condensates }
We define for convenience:
\beq
d_8\equiv  C_8\la O_8\ra.
\label{eq:d8}
\eeq
The contribution of the $D=8$ condensates has been computed in the chiral limit\,\cite{BROAD8}. Higher dimensions operators $(D\geq 10)$ will be omitted as well as contributions due to the estimate of large order uncalculated terms. 

\section{QCD parameters from $e^+e^-\to$ I=1 hadrons data }
We shall determine the QCD  parameters from $e^+e^-\to$ I=1 hadrons data. In so doing, 
we extract the $e^+e^-\to$ I=1 hadrons data component  from   the compilation of the sum of the exclusive decays given by PDG\,\cite{PDG} for the ratio :
\beq
R_{ee}=\frac{\sigma(e^+e^-\to\,{\rm Hadrons}}{\sigma(e^+e^-\to\mu^+\mu^-)}.
\label{eq:Ree}
\eeq
For $e^+e^-\to \pi\pi$, we shall compare the PDG compiled data with the new CMD3 measurement of the pion form factor\,\cite{CMD3}. We take the data below $E_{cm}$=1.875 GeV where they remain quite precise. Then we subdivide the data into three regions. More subdivisions will be done for extracting with a better accuracy the SM-high precion parameters.
\subsection*{\b Warnings}
\d We assume that some issues related to the correlations of data and some other experimental technical subtleties  have been considered in the PDG compilation. 

\d Along this paper, we shall use the Mathematica program FindFit with optimized $\chi^2$ and the program NSolve for our numerical analysis. 

\d To be conservative, we shall fit separately the ensemble of high data points and low data points. Our final result will be the mean of these two extremal values. 
\subsection*{\b Region from $2m_\pi$ to   0.993 GeV}

\begin{figure}[hbt]
\begin{center}
\includegraphics[width=11.cm]{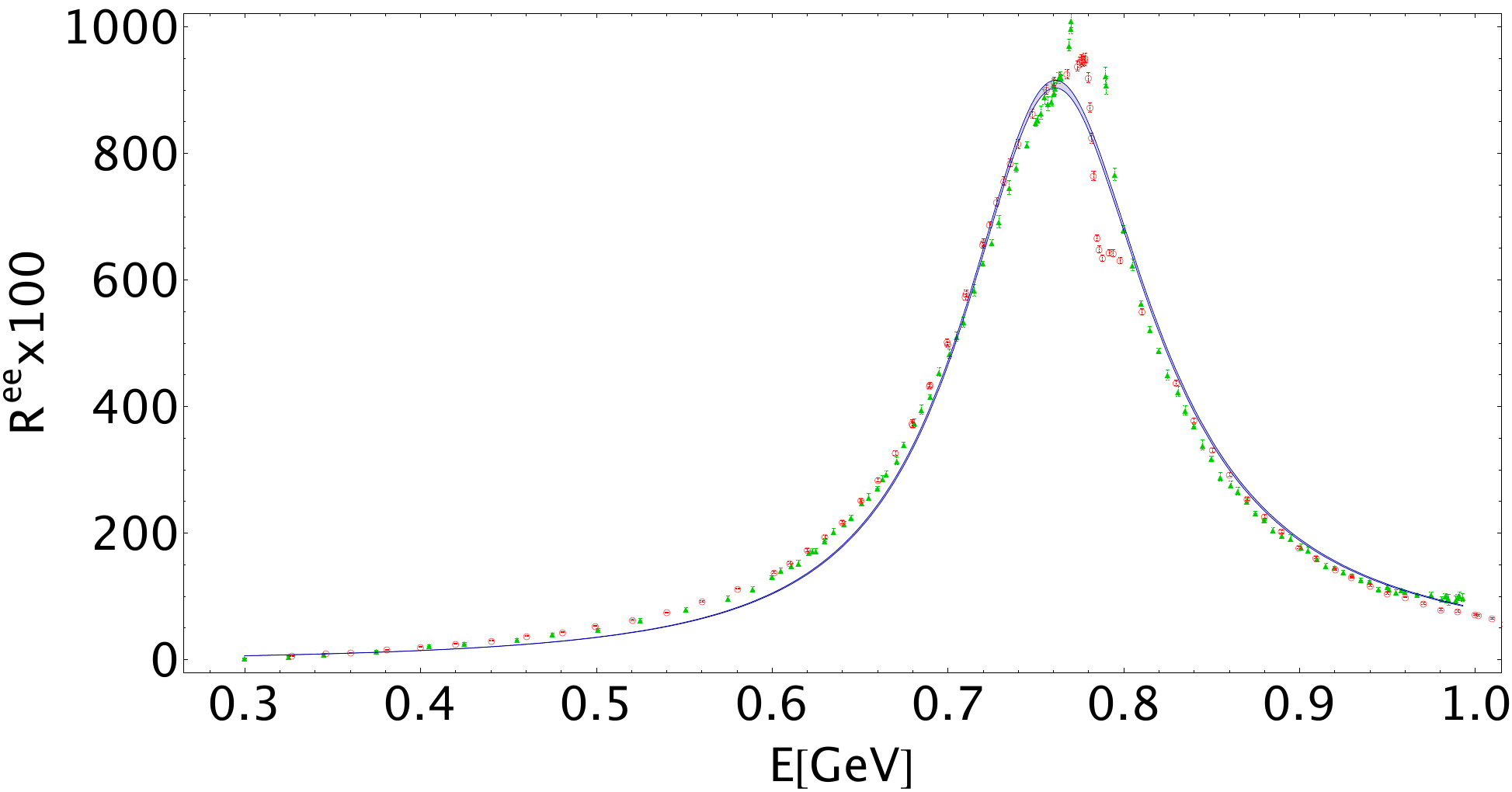}
\caption{\footnotesize  Fit of the data without $\omega$  using a minimal Breit-Wigner parametrization of the PDG\,\cite{PDG} compilation (green triangle) . A comparison with the new CMD3\,\cite{CMD3} data is given (open red circle). . } \label{fig:rho}
\end{center}
\vspace*{-0.5cm}
\end{figure}

We isolate the $I=1$ part by subtracting  the $\omega$ meson contribution and the  shoulder around the peak 
due to the $\omega-\rho$ mixing. Then, we fit the data using the  optimized $\chi^2$ Mathematica program {\it Findfit}. In so doing we fit separately the maximum and minimum values of each data points. To minimize the number of the free parameters, we use as input the value $\Gamma(\rho\to e^+e^-)$ from PDG\,\cite{PDG}. Then, we obtain the parameters:
\bea
{\rm Set}\, 1:  && \Gamma(\rho\to e^+e^-)=(7.03+ 0.04)\,{\rm keV}\lrar  M_{\rho}=755.56\, {\rm MeV} ,~~~~~~~~\Gamma_\rho^{\rm tot}=  132.04\,\rm{MeV}.\nnb\\ 
{\rm Set}\, 2:  && \Gamma(\rho\to e^+e^-)=(7.03-0.04)\,{\rm keV}\lrar  M_{\rho}=755.65\, {\rm MeV} ,~~~~~~~~\Gamma_\rho^{\rm tot}=  131.92\,\rm{MeV},
\eea  
where the mass slightly differs from  $M_\rho=775.5$ MeV quoted by PDG\,\cite{PDG} obtained using a more involved parametrization of the pion form factor and including $\rho-\omega$ mixing. The result corresponds to the small dashed region  shown in Fig.\,\ref{fig:rho} from our conservative fitting procedure. We have not tried to add more parameters on the pion form factor to improve the quality of the fit which is satisfactory for our purpose.  In the section of lepton anomalies where more precise fits are required, we shall improve the fitting procedure by subdividing the regions. We show for a comparison the new data from CMD3 on the pion form factor\,\cite{CMD3}.
The new data are 2-3 times more precise than the one compiled by PDG\,\cite{PDG} but the central values are approximately similar except in the low energy region where the CMD3\,\cite{CMD3} data are slightly higher.  
\subsection*{\b Region from 0.993 to  1.5 GeV}
\begin{figure}[hbt]
\begin{center}
\includegraphics[width=11.cm]{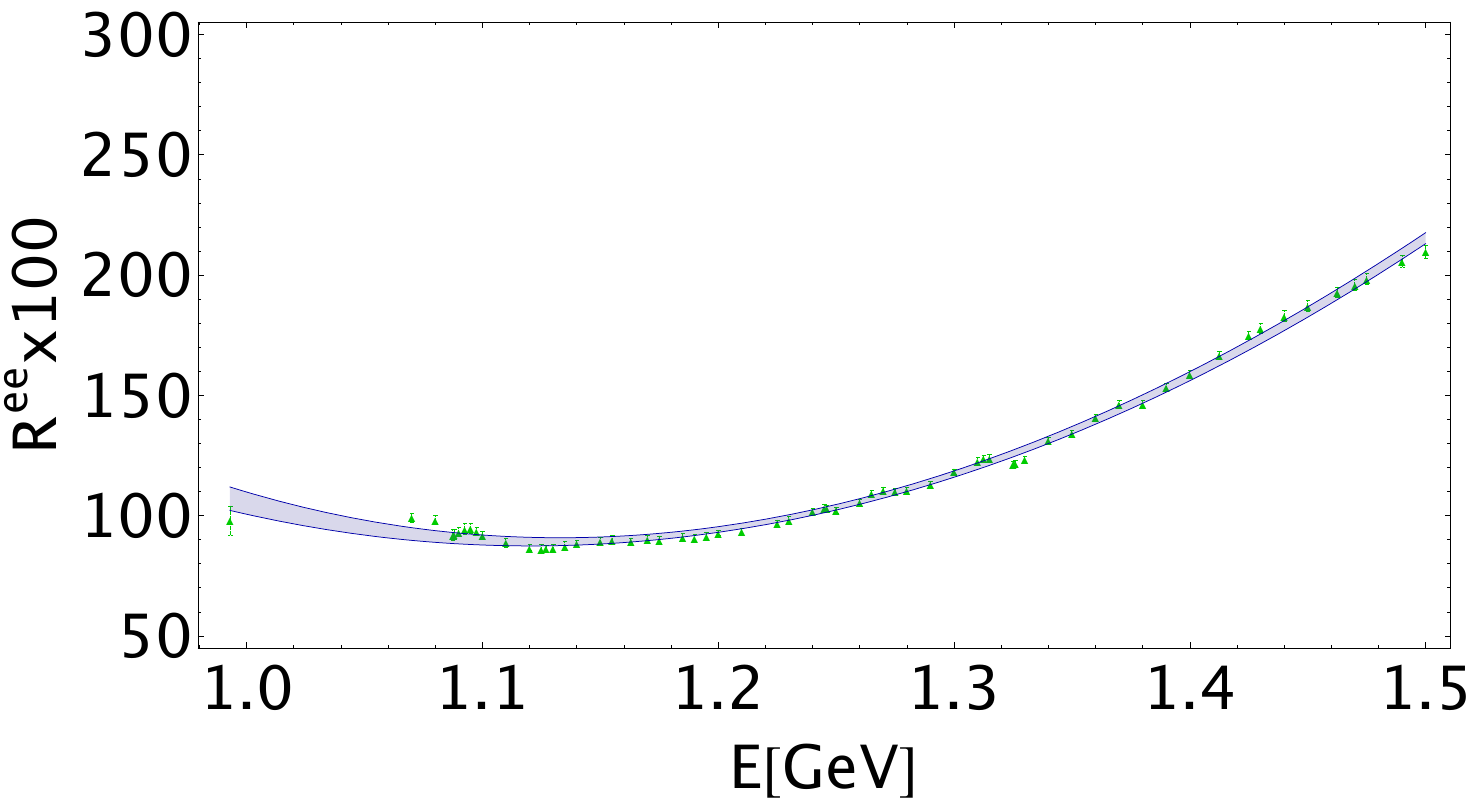}
\caption{\footnotesize  Fit of the data using a cubic polynomial  interpolation  formula.} \label{fig:3pi}
\end{center}
\vspace*{-0.5cm}
\end{figure}

We fit the data using a simple interpolation program with polynomials. We substract the $3\pi$ backgrounds by using the $SU(2)$ relation between the isoscalar and isovector states (a suppression 1/9 factor). 
We neglect the $\bar KK$ contributions from isoscalar sources which, in addition to the $SU(3)$ suppression factor is also suppressed by phase space. The fit is shown in Fig.\ref{fig:3pi}.

\subsection*{\b Region from 1.5 to   1.875 GeV}
\begin{figure}[hbt]
\begin{center}
\includegraphics[width=11.cm]{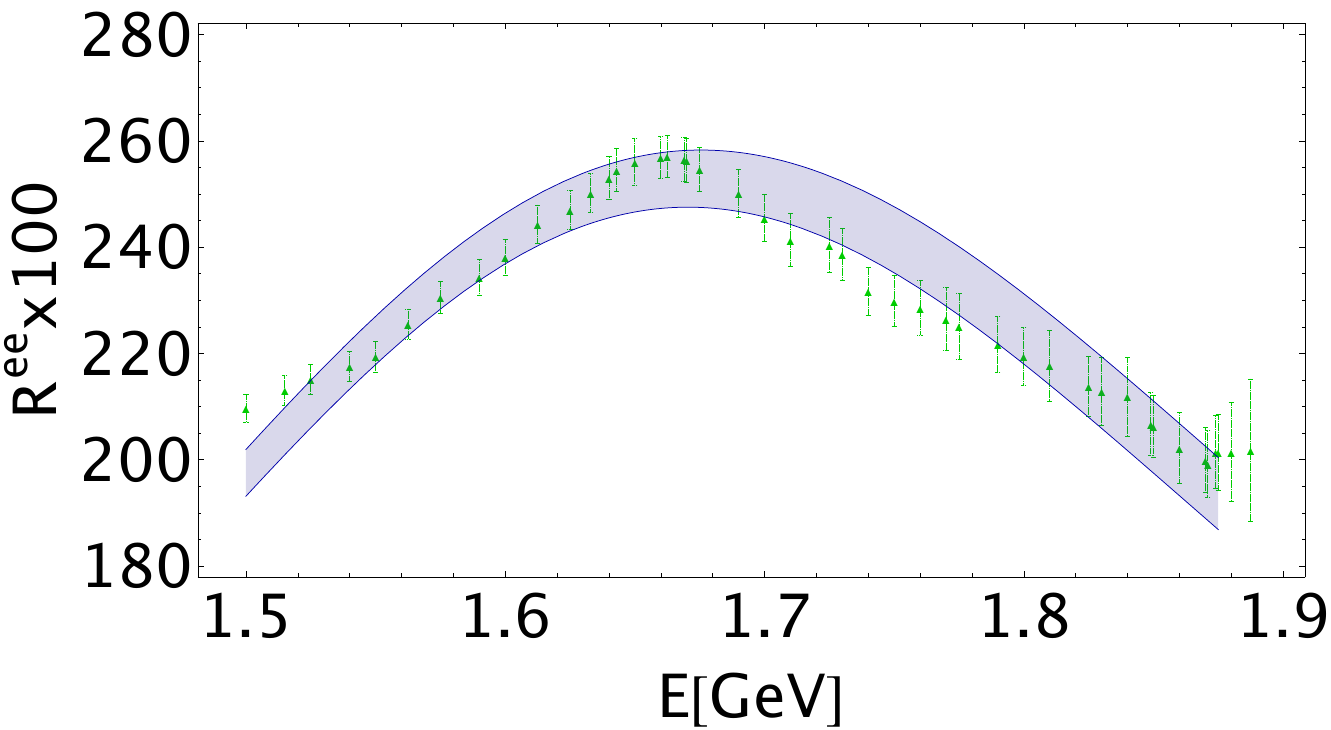}
\caption{\footnotesize  Fit of the data using a minimal Breit-Wigner parametrization.} \label{fig:rhoprime}.
\end{center}
\vspace*{-0.5cm}
\end{figure}

 We use a Breit-Wigner fit for a $\rho'$ meson. We obtain :
\beq
M_{\rho'}=1.6\,{\rm GeV}\,\,\, {\rm and}\,\,\,\Gamma(\rho'\to e^+e^-) = [10.5\,({\rm resp.}\, 9.73)]~{\rm keV}~~~~~~~~\Gamma_{\rho'}^{\rm tot}=  [720\,({\rm resp.}\,694)]\,\rm{MeV},
\eeq
from the high (resp. low) data points. The fit is shown in Fig.\,\ref{fig:rhoprime}.

\vspace*{-0.3cm}
\section{The (inverse) Laplace sum rules}
\vspace*{-0.20cm}
The sum rule reads\,\cite{SVZa,BELLa,BERTa,SNR}:
\bea
{\cal L}^c_0(\tau,\mu)&\equiv&\lim_ {\begin{tabular}{c}
$Q^2,n\to\infty$ \\ $n/Q^2\equiv\tau$
\end{tabular}}
\frac{(-Q^2)^n}{(n-1)!}\frac{\partial^n \Pi}{ ( \partial Q^2)^n}
=\int_{t>}^{t_c}dt~e^{-t\tau}  R_{ee}^{I=1}(t,\mu) 
\nnb\\
 {\cal R}^c_{10}(\tau)&\equiv&\frac{{\cal L}^c_{1}} {{\cal L}^c_0}= \frac{\int_{t>}^{t_c}dt~e^{-t\tau}t\, R_{ee}^{I=1}(t,\mu) }   {\int_{t>}^{t_c}dt~e^{-t\tau} R_{ee}^{I=1}(t,\mu) },~~~~
\label{eq:lsr}
\eea
where $\tau$ is the LSR variable, $t>$   is the hadronic threshold.  Here $t_c$ is  the threshold of the ``QCD continuum" which parametrizes, from the discontinuity of the Feynman diagrams, the spectral function  ${\rm Im}\,\Pi_H(t,m_q^2,\mu^2)$.  $m_q$ is the quark mass and $\mu$ is an arbitrary subtraction point. The spectral function is related  through the optical theorem to the isovector part of the ratio $R^{I=1}_{ee}$ defined as in Eq.\,\ref{eq:Ree} as:
\beq
R^{I=1}_{ee}=\ga\frac{3}{2}\dr 8\pi\, {\rm Im} \Pi_H(t).
\eeq
Due to its superficial convergence, the Laplace sum rule obeys the homogeneous renormalization group equation (RGE)\,\cite{SNR}
\beq
\Big{\{} -\tau\frac{\partial}{\partial\tau}+\beta(\alpha_s)\frac{\partial}{\partial\alpha_s} -\sum_i[1+\gamma(\alpha_s)x_i\frac{\partial}{\partial x_i}  \Big{\}}
{\cal{L}}_n(\tau),
\eeq
with $x_i\equiv m_i(\mu)/\mu$ and $\beta(\alpha_s), \gamma(\alpha_s)$ are the $\beta$ function associated to the renormalization of $\alpha_s$ and anomalous quark mass renormalization in the $\overline{MS}$ scheme. The natural solution to this RGE is the choice $\mu^2=1/\tau$. The QCD expression of the  first moment and the ratio of moments read:
\bea
{\cal L}^{qcd}_0(\tau)&=&\tau^{-1}\frac{3}{2}\Bigg{[}1+a_s+a_s^2( 1.6398-\frac{\beta_1}{2}\gamma_E)+{\cal O}(\alpha_s^3)+ d_2\tau+d_4\tau^2+\frac{d_6}{2}\tau^3+\frac{d_8}{6}\tau^4\Bigg{]},\nnb\\
{\cal R}^{qcd}_{10}(\tau)&=&\tau^{-1}\Bigg{[} 1+\beta_1 a_s^2+{\cal O}(\alpha_s^3)-d_2\tau-2d_4\tau^2-\frac{3}{2}d_6\tau^3-\ga\frac{2}{3}d_8-d_4^2\dr\tau^4\Bigg{]},
\eea
where terms due to the Laplace operator in the PT corrections are explicitly shown. 
Thanks to the exponential weight, the previous sum rule improvements enhance the low energy contribution to the spectral integral which is accessible experimentally. The PT corrections to the ratio of moments ${\cal R}_{10}$ start to order $\alpha_s^2$ which renders it a good observable for extracting the non-perturbative condensates.  Using the $e^+e^-\to $ Hadrons data, Ref.\,\cite{LNT} have used ${\cal R}_{10}$
to extract the condensates and obtained (notice the different normalizations):
\beq
d_4=(3.45\pm 0.85)\times 10^{-2}\,{\rm GeV^4},\,\,\,  d_6=(0.12\pm 0.04)\times 10^{-2}\,{\rm GeV^6},\,\,\,  d_8=(2.9\pm 2.4)\,{\rm GeV^8},
\eeq
where the value of $d_8$ is not conclusive while the one of $d_4$ and $d_6$ lead to the values:
\beq
\la\alpha_s G^2\ra= (3.29\pm 0.81)\times 10^{-2}\,{\rm GeV^4}\,\,\,\,\,\, \rho\la\bar\psi\psi\ra^2=(3.5\pm 1.16)\times 10^{-4}\,{\rm GeV^6}.
\label{eq:res1}
\eeq
The value of $\la\alpha_s G^2\ra$ agrees with the original SVZ\cite{SVZa} value of $3.8\times 10^{-2}\,{\rm GeV^4}$ while the value of the four-quark condensate is larger by a factor 3 compared to the factorization estimate in Eq.\,\ref{eq:fac}. However, a more recent analysis using $R_{10}$ and the $\tau$-like decay sum rules for the $I=1$ hadrons data lead to\,\cite{SN95,SNTAU}:
\beq
\la\alpha_s G^2\ra= (7.1\pm 0.7)\times 10^{-2}\,{\rm GeV^4}\,\,\,\, \rho\la\bar\psi\psi\ra^2=(5.8\pm 0.9)\times 10^{-4}\,{\rm GeV^6},\,\,\,  d_8=-(0.38\pm 0.15)\,{\rm GeV^8}.
\label{eq:res2}
\eeq
One should note from Eqs.\,\ref{eq:res1} and \ref{eq:res2} that there is a discrepancy by a factor of about 2 for the estimate of $\la\alpha_s G^2\ra$ and $\rho\la\bar\psi\psi\ra^2$ but both results indicate a huge violation of the four-quark vacuum saturation.

In the following, our aim is to try to clarify the contradictory results from the SVZ-like sum rules and from $\tau$-decay moments using  the example of vector current channel and the lowest moments and their ratio. 

\vspace*{-0.3cm}
\section{Fixing the value of the QCD continuum threshold $t_c$}
\vspace*{-0.20cm}
First, we have to fix the value of the QCD continuum threshold $t_c$ above  which we shall parametrize the spectral function
by its QCD expression given in Eq.\,\ref{eq:pt}. 
\subsection*{\b $t_c$ from the moment  ${\cal L}^{ee}_{0}$}
\begin{figure}[hbt]
\begin{center}
\includegraphics[width=11.cm]{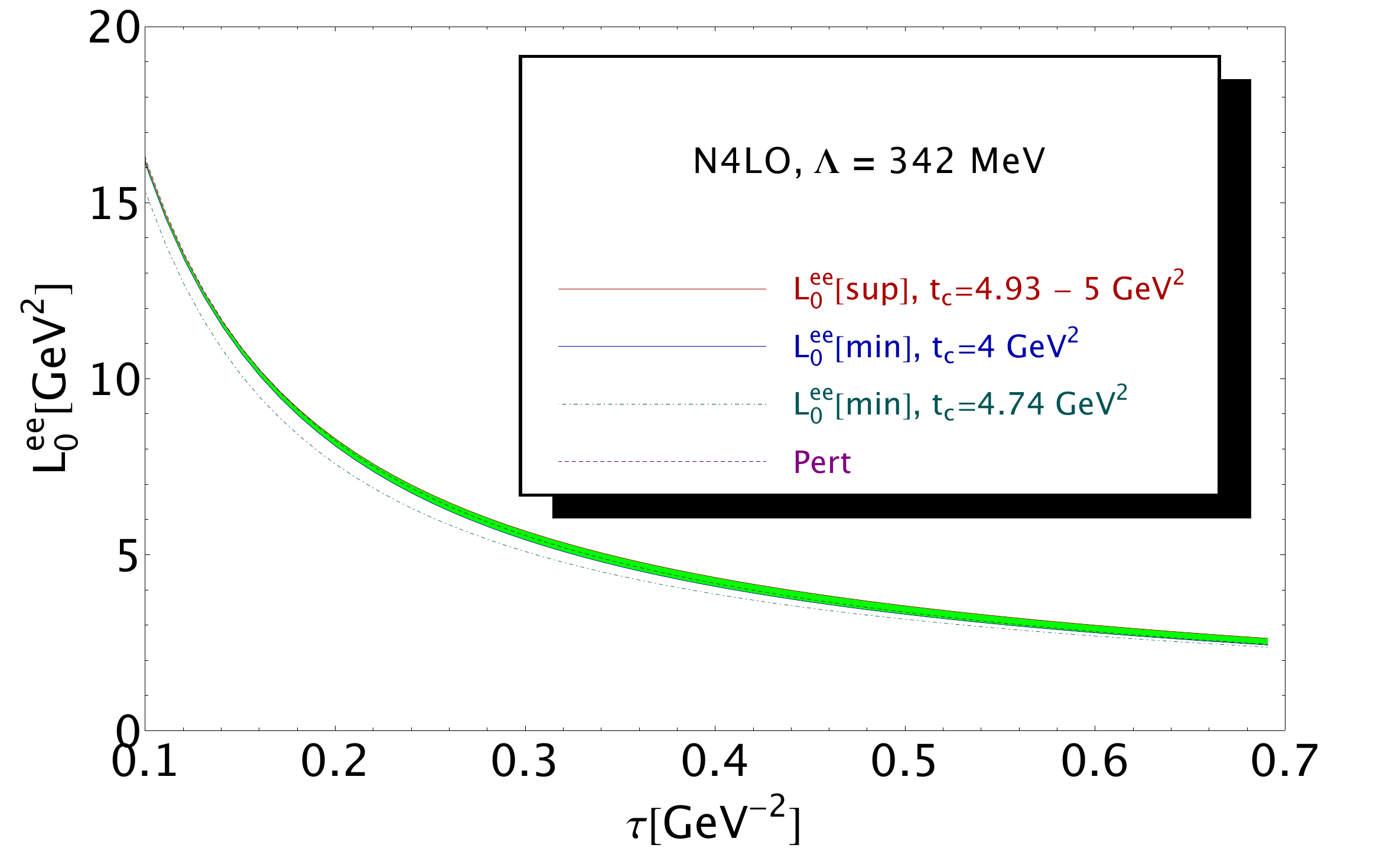}
\caption{\footnotesize  Comparison of the QCD expression of ${\cal L}^{ee}_{0}$ with the data for the obtained value of $t_c$ given in Eq.\, \ref{eq:tc}} \label{fig:L0-tc}.
\end{center}
\vspace*{-0.5cm}
\end{figure}

We request that the data and the perturbative QCD expression of the moment ${\cal L}^{ee}_0$ coincides for the sum rule variable $\tau\leq 0.7$ GeV$^{-2}$ where the non-perturbative effects to the QCD expression are expected to be negligible while the difference between the QCD contribution with the data one without a QCD continuum remains positive.  Using the  Mathematica optimized $\chi^2$ fit program FindFit. We obtain  to order $\alpha_s^4$ accuracy :
\beq
t_c=(4.74\sim 4.93) ~{\rm GeV}^2,
\label{eq:tc1}
\eeq
where $t_c=4.74$ (resp.4.93) GeV$^2$ corresponds to the fit of the low (resp. high) set of values of the data points. The result of the analysis is shown in Fig.\ref{fig:L0-tc}.  We have improved  the fit for the low data points by adjusting manually the value of $t_c$. Then, we arrive at  the  conservative range of values :
\beq
t_c=(4\sim 5) ~{\rm GeV}^2,
\label{eq:tc}
\eeq
which we shall use for the forthcoming analysis. 

\subsection*{\b Comments and comparison}
\begin{figure}[hbt]
\begin{center}
\includegraphics[width=11.cm]{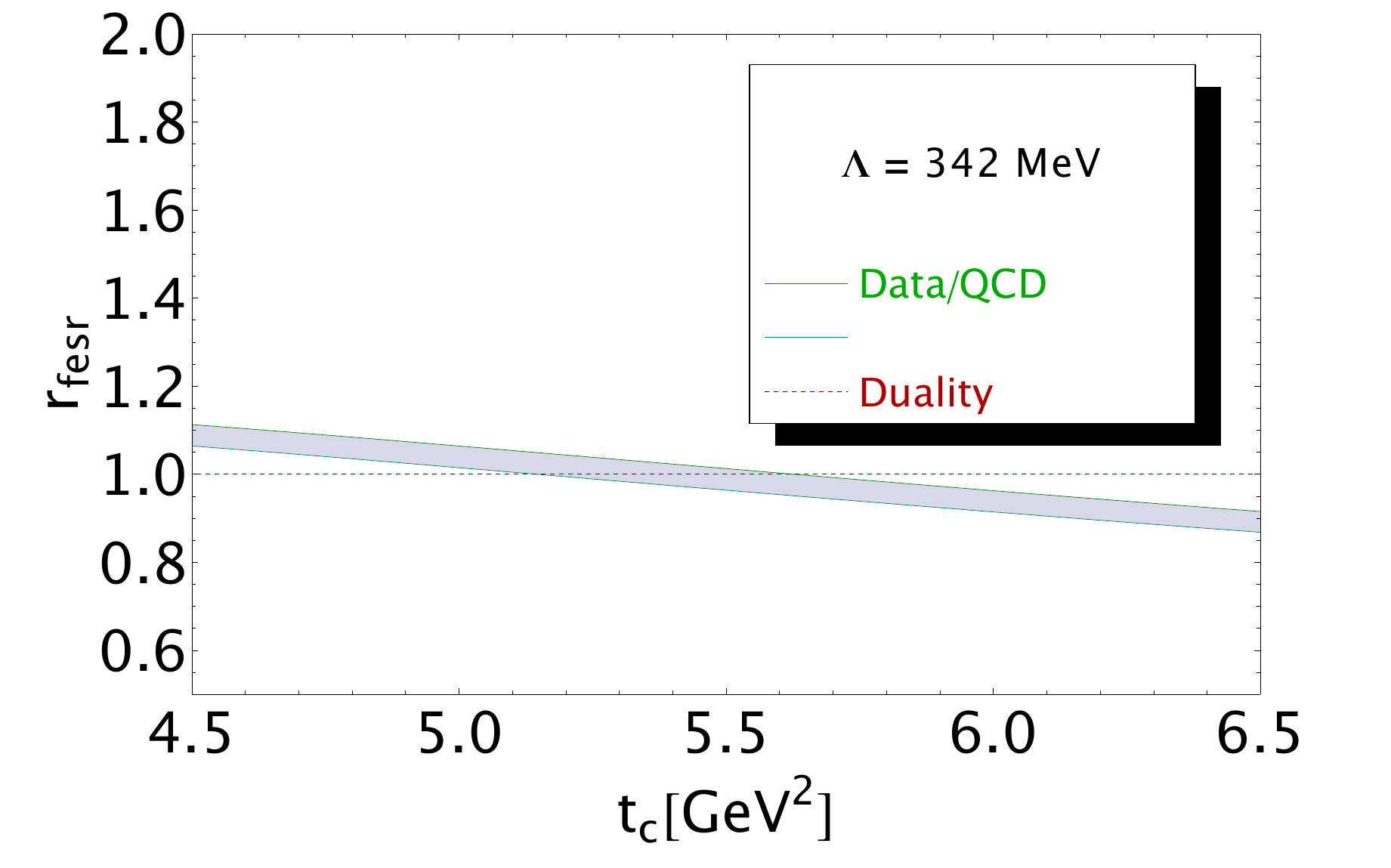}
\caption{\footnotesize  Variation of the ratio in Eq.\,\ref{eq:ratio-fesr} versus $t_c$.}\label{fig:fesr}
\end{center}
\vspace*{-0.5cm}
\end{figure}

It is informative to compare this result with the one required by FESR  for a duality between the experimental and QCD side of the sum rule\,\cite{FESR}:
\beq
F_0(t_c)\equiv \int_0^{t_c} dt R^{I=1}_{ee} (t).=\frac{3}{2}t_c\Bigg{[} 1+a_s(t_c)+a_s^2(t_c)\ga 1.6398-\frac{\beta_1}{2}\dr+\cdots\Bigg{]}
\eeq
where $\cdots$ comes from a numerical integration of the integral at order $\alpha_s^3$. The analysis for the ratio
\beq
R_0^{fesr}\equiv \frac{F_0^{data}} {F_0^{qcd}}
\label{eq:ratio-fesr}
\eeq
 is shown in Fig.\,\ref{fig:fesr} from which one can deduce to order $\alpha_s^3$ the duality region:
\beq
t_c^{fesr} =(5.1\sim 5.6) ~{\rm GeV}^2,
\label{eq:fesr}
\eeq
The value in Eq.\,\ref{eq:fesr} is higher than the value around 2.1 GeV$^2$ obtained in Ref.\,\cite{FESR}. We interpret the origin of this discrepancy as due to the effect of the old data and of its parametrization in the $\rho'$-region.  This value  does not also justify  the choice $t_c=1.55$ GeV$^2$ used  in Ref.\,\cite{BOITO} for the  unpinched moment which is  equivalent to the lowest degree of FESR used here.  
\vspace*{-0.3cm}
\section{QCD  condensates from the ratio ${\cal R}_{10}$ of  Laplace  sum rules moments}
\vspace*{-0.20cm}
We shall extract the condensates from the ratio of moments which has the advantage to be less sensitive to the PT $\alpha_s$ corrections which tends to cancel out in the ratio. 
\subsection*{\b $\la \alpha_sG^2\ra$ and $d_6$ from the ratio of moments ${\cal R}_{10}$ to order $\alpha_s^2$}\label{sec:d46}
\begin{figure}[hbt]
\begin{center}
\includegraphics[width=11.cm]{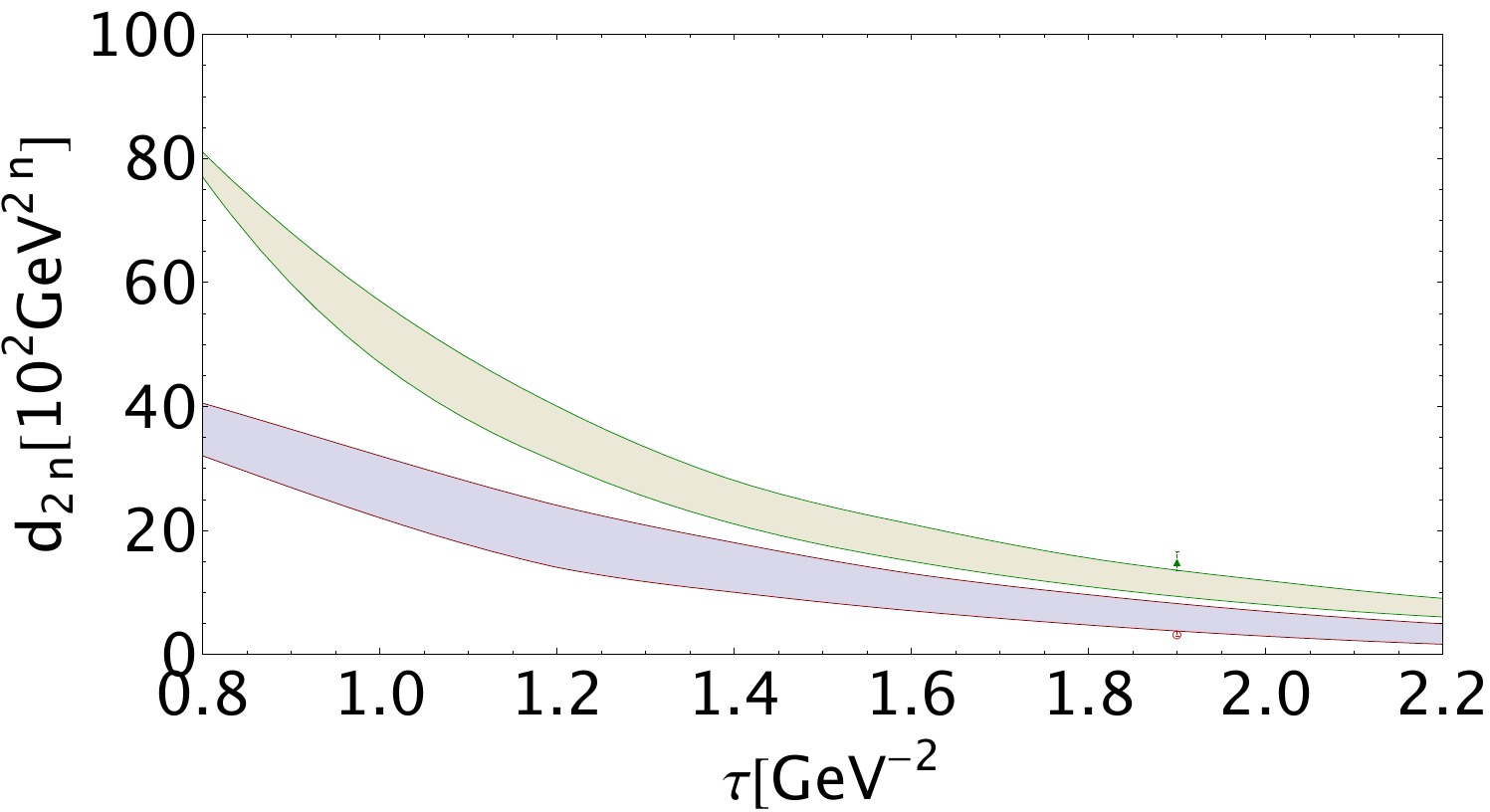}
\caption{\footnotesize  Two-parameter fit of 
$\la \alpha_sG^2\ra$ and $-d_6$ from the ratio of moments ${\cal R}_{10}$.} \label{fig:d46}.
\end{center}
\vspace*{-0.5cm}
\end{figure}

We attempt to extract the gluon condensate $\la \alpha_sG^2\ra$ and the $D=6$ four-quark condensate from the ratio of moments to order $\alpha_s^2$ by neglecting the contributions of $D\geq 8$ ones. The analysis is shown in Fig.\,\ref{fig:d46} versus the sum rule variable $\tau$. The green curves delimit the value of $\la \alpha_sG^2\ra$ and the red ones the value of $d_6$. 
We notice that there is no stability in $\tau$ such that the analysis is unconclusive. The results obtained from a fit by LNT in Ref.\,\cite{LNT} are shown by the full triangle ($\la \alpha_sG^2\ra$) and open circle ($d_6$).

\subsection*{\b $d_6$ and $d_8$ from the ratio of moments ${\cal R}_{10}$ to order $\alpha_s^2$}
 We  impose more constraints on  the parameters by using as input the value of the gluon condensate given in Eq.\,\ref{eq:asg2} from heavy quark sum rules\,\footnote{A 3-parameter fit does not lead to a conclusive result.}.  As mentioned in the introduction, we shall truncate the OPE at $D=8$ and assume that the $D=8$ condensate is an effective condensate which absorbs into it all higher dimension ones. One also should note that looking for a combination of these high-dimension condensates to have a physical meaning is a difficult task as they mix under renormalization\,\cite{SNTARRACH}. 

\begin{figure}[hbt]
\begin{center}
\includegraphics[width=11.cm]{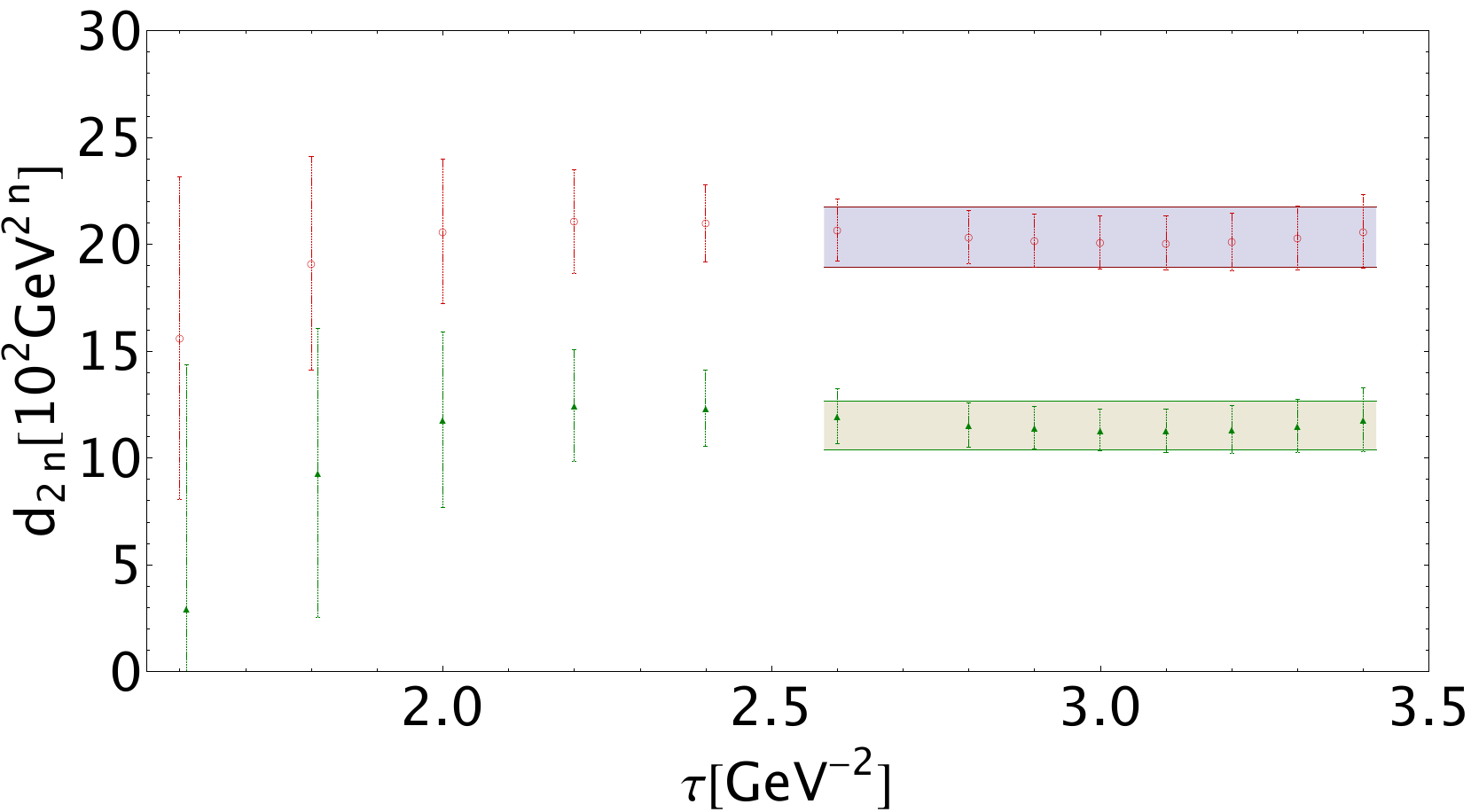}
\caption{\footnotesize  Two-parameter fit of $-d_6$ and $+d_8$  to order $\alpha_s^2$ for a given value of $\la\alpha_s G^2\ra$ from Eq.\,\ref{eq:asg2}.} \label{fig:d68}.
\end{center}
\vspace*{-0.5cm}
\end{figure}

Given the previous value of $t_c$ in Eq.\,\ref{eq:tc}, we study the effect of the Laplace sum rule variable $\tau$ on the result. This is shown in Fig.\,\ref{fig:d68}.  
One can observe a $\tau$-stability  at $\tau\simeq 2.6$  to 3.4  GeV$^{-2}$ which corresponds to the values:
\beq
d_6 =  -(20.1\pm1.3)\times 10^{-2}\,{\rm GeV^6},\,\,\,\,\,\,\, d_8= (11.3\pm 1.1)\times 10^{-2}\,{\rm GeV^8}
\label{eq:d68-min}
\eeq 
at order $\alpha_s^2$. 
As a final value, we consider  the average of different results obtained  in the region $\tau\simeq 2.6$ to 3.4 GeV$^{-2}$ around this minimum. We deduce (shaded region):
\beq
d_6 =  -(20.3\pm1.4)\times 10^{-2}\,{\rm GeV^6},\,\,\,\,\,\,\, d_8= (11.5\pm 1.2)\times 10^{-2}\,{\rm GeV^8}.
\label{eq:res-d68-as2}
\eeq 
\subsection*{\b QCD PT corrections to order $\alpha_s^4$}
One can include higher order terms by using the expression of ${\cal L}_0$ to order $\alpha_s^4$\,\cite{KAHN2}
:
\beq
{\cal L}^{PT}_0(\tau)= \tau^{-1}\Big{[} 1+a_s+2.93856\,a_s^2+ 6.2985\,a_s^3 + 22.2233\,a_s^4\Big{]},
\eeq
and taking its derivative in $\tau$ to get ${\cal L}_1(\tau)$ and then their ratio ${\cal R}_{10}(\tau)$. 
We use the expression of $\alpha_s$ to order $\alpha_s^3$ given e.g. in Eq. 11.66 of \cite{SNB1}:
\beq
a_s(\tau)=a_s^{(0)}\Bigg{\{} 1-a_s^{(0)}\frac{\beta_2}{\beta_1}l_\tau+\ga a_s^{(0)}\dr^2\Bigg{[} \log^2(l_\tau)-\ga\frac{\beta_2}{\beta_1}\dr^2\, \log(l_\tau)-\ga\frac{\beta_2}{\beta_1}\dr^2+\frac{\beta_3}{\beta_1}\Bigg{]}+{\cal O}(a_s^3)\Bigg{\}},
\eeq
where :   $l_\tau\equiv -\log \tau\,\Lambda^2$, $a_s^{(0)}\equiv 2/(\beta_1 l_\tau)$ and, for three flavours:
\beq
\beta_1=-9/2\, ,\,\,\,\,\,\,\,\,\,\,\,\,\,\,\,\,\,\,\,\,\,\,\,\,\,\,\,\, \beta_2=-8\,   ,\,\,\,\,\,\,\,\,\,\,\,\,\,\,\,\,\,\,\,\,\,\,\,\,\,\,\,  \beta_3=-20.1198~.
\eeq
\begin{figure}[hbt]
\begin{center}
\includegraphics[width=11.cm]{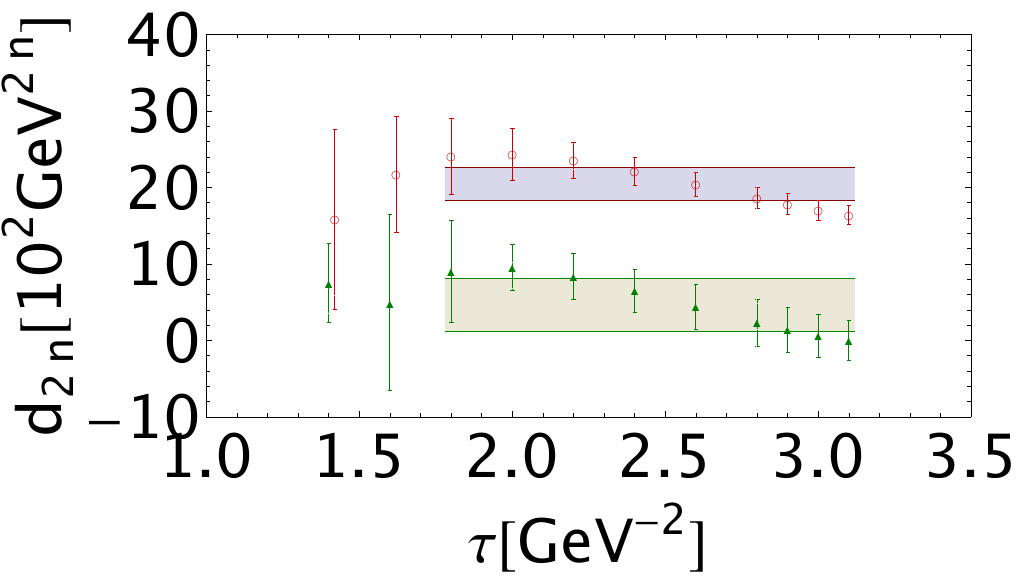}
\caption{\footnotesize  Two-parameter fit of $-d_6$ and $+d_8$ to order $\alpha_s^4$  for a given value of $\la\alpha_s G^2\ra$ from Eq.\,\ref{eq:asg2}.} 
\label{fig:d68-as4}.
\end{center}
\vspace*{-0.5cm}
\end{figure} 
\begin{figure}[hbt]
\begin{center}
\includegraphics[width=11.cm]{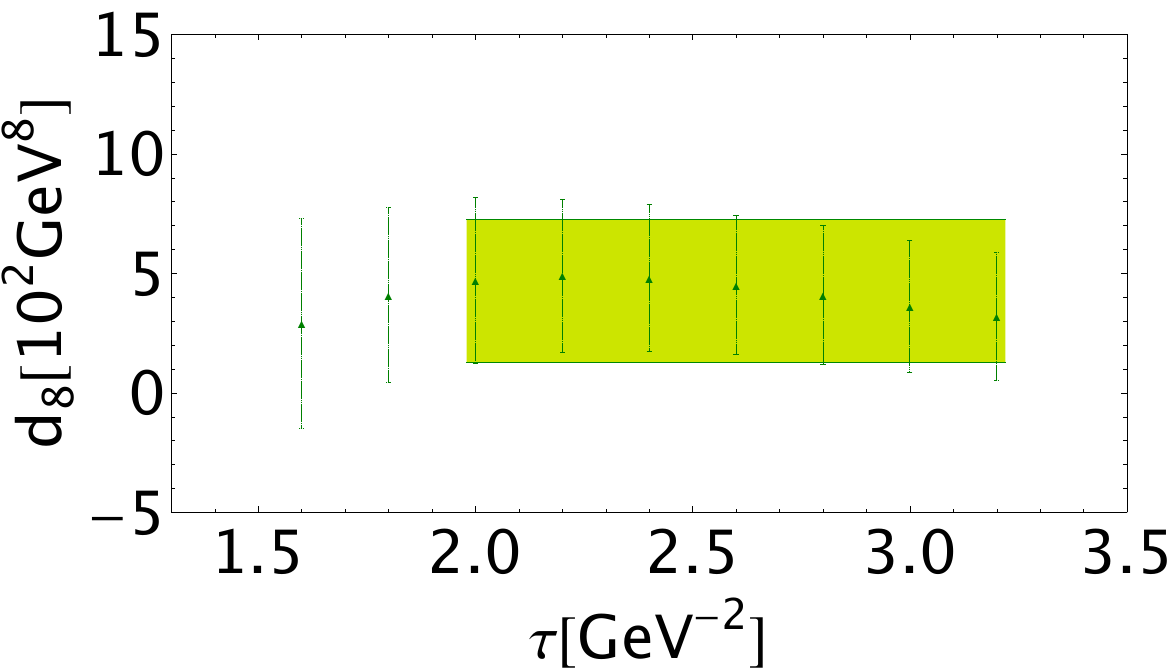}
\caption{\footnotesize  One-parameter fit of $d_8$ to order $\alpha_s^4$  for a given value of $\la\alpha_s G^2\ra$ from Eq.\,\ref{eq:asg2} and $d_6$ from Eq.\,\ref{eq:res-d68}.} 
\label{fig:d8}.
\end{center}
\vspace*{-0.5cm}
\end{figure} 
\subsection*{ \b Extraction of $d_6$ and $d_8$ to order $\alpha_s^4$}
 We show in Fig.\,\ref{fig:d68-as4} the results from a two-parameter fit to order $\alpha_s^4$. By comparing Figs.\,\ref{fig:d68} and \ref{fig:d68-as4}, one can notice that the $\alpha_s^4$-term destabilizes the $\tau$-behaviour of $d_6$ and $d_8$.  A least square fit of the points between 1.8 and 3.2 GeV$^{-2}$ where the determinations are more precise gives:
\beq
d_6 =  -(20.5\pm2.2)\times 10^{-2}\,{\rm GeV^6},\,\,\,\,\,\,\, d_8= (4.7\pm 3.5)\times 10^{-2}\,{\rm GeV^8}.
\label{eq:res-d68}
\eeq 
The value of $d_6$ remains unchanged from $\alpha_s^2$ to $\alpha_s^4$ which is not the case of $d_8$ which has decreased by about a factor 2.5\,! Then, we try to improve the determination of $d_8$ by using $d_6$ as input in addition to the previous input parameters. The analysis shown in Fig.\ref{fig:d8} indicates a much better stability though we do not (unfortunately) gain much on the accuracy. We obtain:
\beq
d_8= (4.3\pm 3.0)\times 10^{-2}\,{\rm GeV^8}.
\label{eq:d8-res}
\eeq
which we consider as a final result for $d_8$. However, this low value of $d_8$ compared to the one obtained at order $\alpha_s^2$ in Eq.\,\ref{eq:res-d68-as2} is a good news for QCD sum rules users. It (a posteriori) justifies the neglect of high-dimension condensates in the analysis which gives a good description of different hadron parameters when only the dimension $D\leq 6$ condensates are retained in the OPE. 

From the value of $d_6$ in Eq.\,\ref{eq:res-d68}, one can deduce the value of  the four-quark condensate determined at order $\alpha_s^4$:
\beq
 \rho\la\bar\psi\psi\ra^2=(5.98\pm 0.64)\times 10^{-4}\,{\rm GeV^6},
 \label{eq:res-4q}
 \eeq
 which (agreably) confirms the value in Eq.\,\ref{eq:psi2} from heavy quarkonia and the average of different determinations quoted in\,\cite{SNparam}. 
\subsection*{ \b  A second attempt to extract $\la\alpha_s G^2\ra$ to order $\alpha_s^4$ }
\subsection*{\hspace*{0.5cm} \d $\la\alpha_s G^2\ra$  and $d_6$ from a two-parameter fit}
\begin{figure}[hbt]
\begin{center}
\includegraphics[width=11.cm]{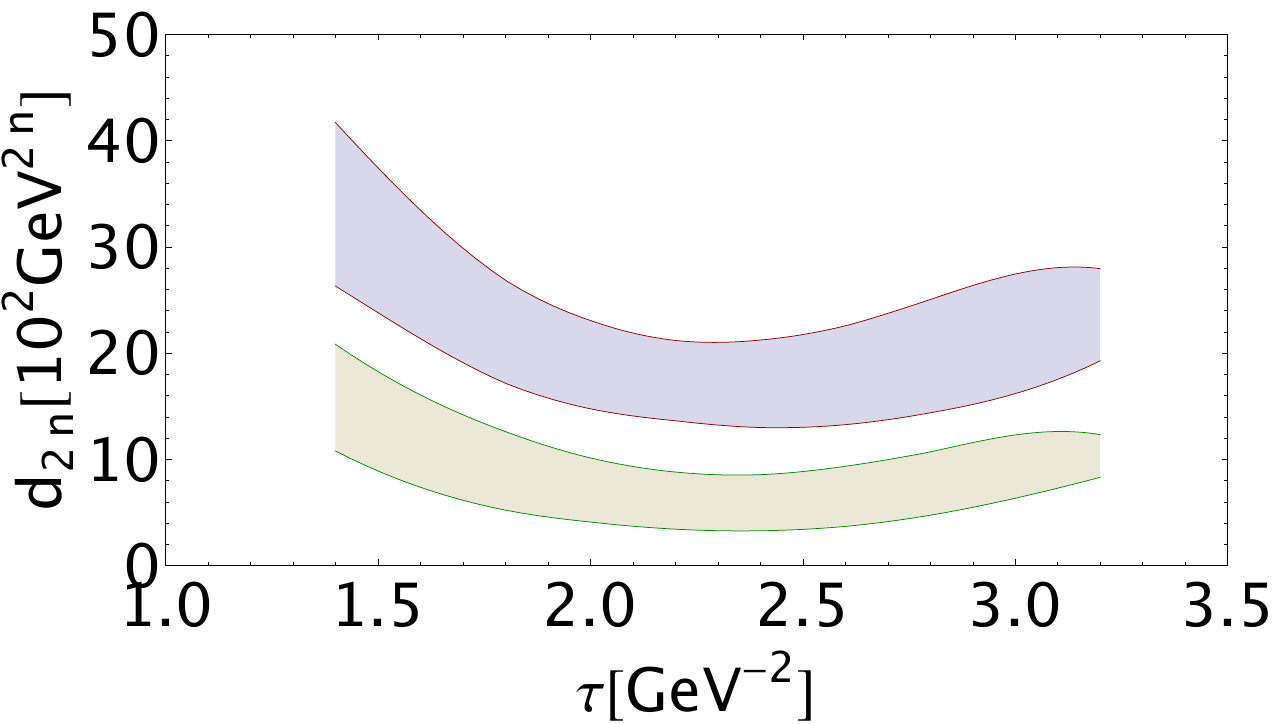}
\caption{\footnotesize  Two-parameter fit of  $\la\alpha_s G^2\ra$ (lower curves) and $d_6$ (upper curves) to order $\alpha_s^4$.} 
\label{fig:d64-as4}.
\end{center}
\vspace*{-0.5cm}
\end{figure}

First, we repeat the analysis done previously  and neglect the $d=8$ condensates. One can notice (see Fig.\,\ref{fig:d64-as4}) that the presence of the $\alpha_s^4$ term leads in this case to $\tau$-stability (minimum) at $\tau\simeq 2.4$ GeV$^{-2}$  from which we extract the optimal values:
\beq
\la\alpha_s G^2\ra =  (5.9\pm 2.6)\times 10^{-2}\,{\rm GeV^4},\,\,\,\,\,\,\, d_6= -(17.1\pm 4.1)\times 10^{-2}\,{\rm GeV^6},
\label{eq:d46}
\eeq 
where the main error is due to the data fitting procedure.
These values agree with the determination of  $\la\alpha_s G^2\ra$ from heavy quarkonia in Eq.\,\ref{eq:asg2} and with the one of $d_6$ in Eq.\,\ref{eq:res-d68} but inaccurate. 
\subsection*{\hspace*{0.5cm} \d $\la\alpha_s G^2\ra$  from a one-parameter fit}
To improve the determination of $\la\alpha_s G^2\ra$, we use as input the value of $d_6$ in Eq.\,\ref{eq:res-d68} and include $d_8$ from Eq.\,\ref{eq:d8-res}. The analysis is shown in Fig.\,ref{fig:d4}. We deduce at the stability point $\tau\simeq 1.5$ GeV$^{-2}$:
\beq
\la\alpha_s G^2\ra =  (6.12\pm 0.61)\times 10^{-2}\,{\rm GeV^4},
\label{eq:ag2}
\eeq
which improves the result in Eq.\,\ref{eq:d46}.  Combining this value with the one from heavy quarkonia in Eq.\,\ref{eq:asg2}, we deduce the average:
\beq
\la\alpha_s G^2\ra =  (6.40\pm 0.30)\times 10^{-2}\,{\rm GeV^4},
\eeq

\begin{figure}[hbt]
\begin{center}
\includegraphics[width=11.cm]{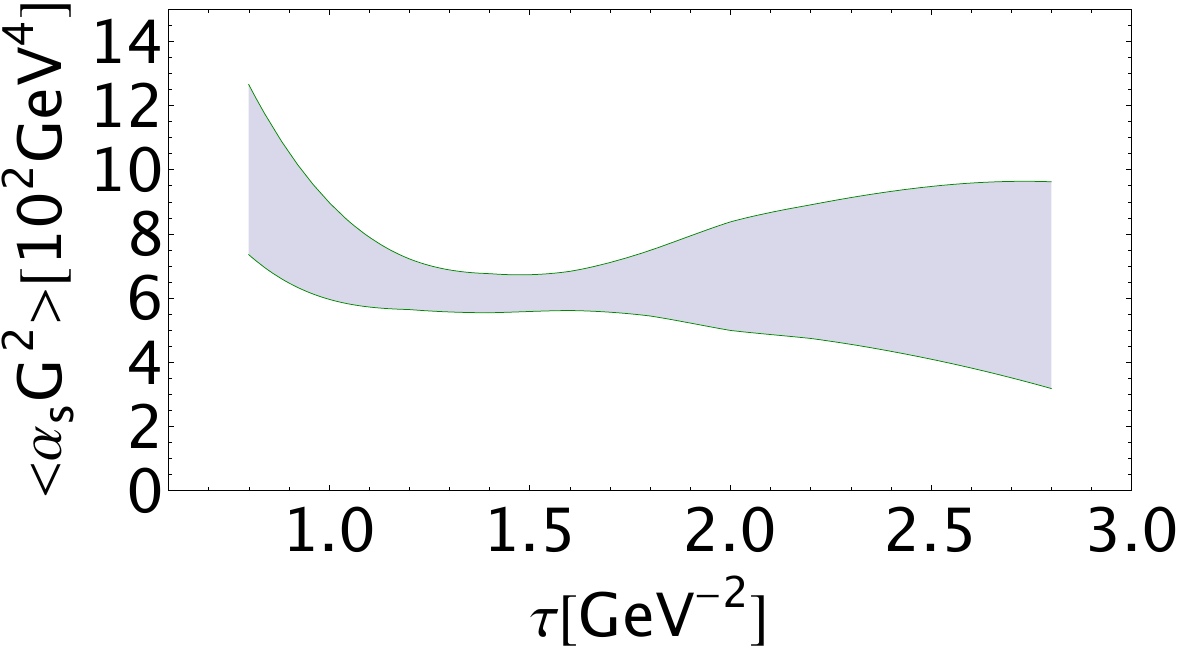}
\caption{\footnotesize  One-parameter fit of  $\la\alpha_s G^2\ra$ to order $\alpha_s^4$ using $d_6$ in Eq.\,\ref{eq:res-d68} and $d_8$ from Eq.\,\ref{eq:d8}.} 
\label{fig:d4}.
\end{center}
\vspace*{-0.5cm}
\end{figure} 

\subsection*{\b Comparison with LNT\,\cite{LNT}}
We consider this result as an improvement of  the one of LNT in Ref.\,\cite{LNT} (notice a slightly different normalization):
\beq
d_6\vert_{lnt} =  -(12.7\pm 4.2)\times 10^{-2}\,{\rm GeV^6},\,\,\,\,\,\,\,\,\,\,\,\,\,\,\,\,\,\,\,\,\,\,\,\,\,\,\,\, d_8\vert_{lnt}= (27\pm 24)\times 10^{-2}\,{\rm GeV^8}
\eeq
 where the same ratio of moments has been used  with a different strategy and data. The lower value of $d_6$ is correlated to the low value of $\la\alpha_s G^2\ra$ 
 obtained by LNT (see Fig.\ref{fig:d46}). 
 
\subsection*{\b Comparison with FESR\,\cite{FESR}}
 It is informative to compare this result with the one from FESR using $e^+e^-\to$ hadrons data:
  \beq
 d_6=\int_0^{t_c} dt \, t^2  R^{I=1}_{ee}(t)-\ga\frac{3}{2}\dr\frac{t_c^3}{3}\Bigg{[} 1+a_s+a_s^2\ga 1.6398-\frac{\beta_1}{6}\dr+\cdots\Bigg{]},
  \label{eq:d6fesr}
  \eeq
  which is interesting as it can disentangle the contributions of the condensates having a given dimension. One obtains\,\cite{FESR}:
 \beq
 d_6\vert_{fesr} = -(0.33-0.55)~{\rm GeV^6},
 \eeq
 where the order of magnitude and the sign agree with the previous estimate. However, we have attempted to rederive the FESR result and noticed that it is quite sensitive to the parametrization of the data at high-energy and to the corresponding value of $t_c$ where $d_6$ behaves as $t_c^3$ in Eq.\,\ref{eq:fesr} such that our analysis has lead to an unconclusive result. We should notice that the FESR needs a fine tuning for  extracting the small $d_6$ number from the difference of two big quantities.  Similar comments apply for the extraction of the gluon condensate from the FESR:
 \beq
 d_4=\ga\frac{3}{2}\dr\frac{t_c^2}{2}\Bigg{[} 1+a_s+a_s^2\ga 1.6398-\frac{\beta_1}{4}\dr+\cdots\Bigg{]}-\int_0^{t_c} dt \, t \, R^{I=1}_{ee}(t)-.
  \eeq
 It has been found that\,\cite{FESR}\,:
 \beq
 \la\alpha_s G^2\ra\vert_{fesr}\simeq (7\sim 18)\times 10^{-2}~{\rm GeV}^4.
 \eeq 
  Despite its inaccuracy, the range of value of the gluon condensate agrees with the one from different  channels in Eq.\,\ref{eq:asg2} but disfavours the negative value from $\tau$-decay moments using CI \,\cite{DAVIER,PICH1}. The inaccuracy of the FESR result also comes from the $t_c^2$ dependence for its extraction.
  \subsection*{\b Comparison with $\tau$-decay moments}
   Different groups\,\cite{BOITO,ALEPH,OPAL,PICH1,PICH2,DAVIER} have  worked  with a set of moments introduced  in Ref.\,\cite{LEDI}  and its variants which are a generalization of the original lowest $\tau$-decay rate moment used by BNP\,\cite{BNP2,BNP}.  
    Fitting   three / four free  parameters ($\alpha_s, d_4,d_6,\cdots$), the authors obtain results which vary in a large range. We compare our results with the ones  from the vector component analysis. 
  \subsection*{\d Gluon condensate $\la\alpha_s G^2\ra$ }
   Its extraction is affected by the way of truncating the QCD series. Within the standard OPE, it can become negative for CI. Its values from different groups given in Table\,\ref{tab:g2} vary in a large range which may indicate that $\tau$-decay like-moments  are not a good tools for extracting $\la \alpha_s G^2\ra$.  We take the values from Table 1 of Ref.\,\cite{PICH1} which is consistent with the fact that we neglect the $D\geq 10$ condensate contributions.
   
   {\normalsize
\begin{table}[hbt]
\setlength{\tabcolsep}{1.2pc}
  \begin{center}
    {
  \begin{tabular}{lllll}

&\\
\hline
\hline
$\la\alpha_s G^2\ra$&$-d_6$&$d_8$&PT series & Refs.\\
 \hline 
$0.67\pm 0.89$&$15.2\pm 2.2$&$22.3\pm 2.5$&    {\rm FO}&{\rm ALEPH}\,\cite{ALEPH}\\
$5.34\pm 3.64$&$14.2\pm 3.5$&$21.3\pm 2.5$&            {\rm FO}&{\rm OPAL}\,\cite{OPAL}\\
 $0.8^{+0.7}_{-1.4}$&$32^{+8}_{-5}$&$50^{+4}_{-7}$&{\rm FO}&\,\cite{PICH1}\\
$0.31\pm 2.45$&$13.5\pm 1.8$&$20.0\pm 1.6$&       {\rm CI}&{\rm OPAL}\,\cite{OPAL}\\
$-1.57\pm0.94$&$14.7\pm 1.1$&$20.4\pm 1.3$&{\rm CI}&\,{\rm ALEPH}\,\cite{DAVIER}\\
$-0.8^{+0.7}_{-0.7}$&$35\pm 3$&$49 ^{+4}_{-5}$&  {\rm CI}&\,\cite{PICH1}\\
   \hline\hline
\end{tabular}}
 \caption{ Values of the QCD condensates of dimension $D=2n$ in units of $10^{-2}$ GeV$^{2n}$ from $\tau$-like decays.  }\label{tab:g2} 
 \end{center}
\end{table}
} 
  \subsection*{\d $d_6$ and $d_8$  condensates }
Their determinations from $\tau$-like moments are compiled in Table\,\ref{tab:g2}. One can notice that, unlike the gluon condensate, 
they are less affected by the truncation of the PT series. Our values given in Eq.\,\ref{eq:res-d68} are in the range of these values 
though the size of our $d_8$ is about half. 
  \subsection*{ \b Correlated values of $d_6$ and $d_8$ versus $\la\alpha_s G^2\ra$}
\begin{figure}[hbt]
\begin{center}
\includegraphics[width=11.cm]{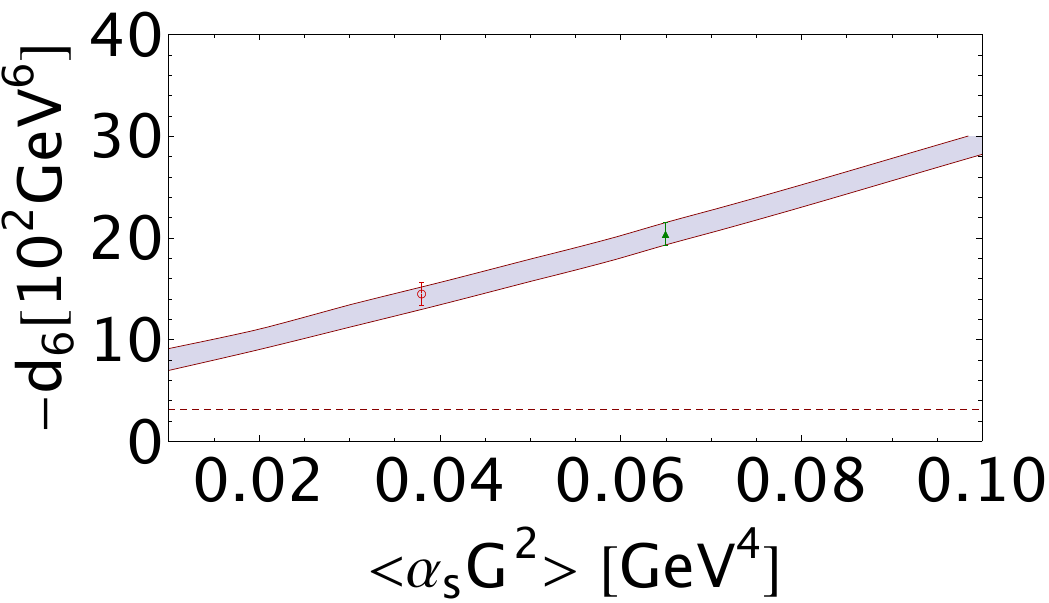}
\caption{\footnotesize  Correlated values of $d_6$ and $d_8$ versus $\la\alpha_s G^2\ra.$ The dashed horizontal line is the value of $d_6$ estimated from factorization of the four-quark condensate. The red (resp. oliva) point corresponds to the value of $\la\alpha_s G^2\ra$ given by SVZ and by Eq.\,\ref{eq:asg2}. } \label{fig:d68-g2}.
\end{center}
\vspace*{-0.5cm}
\end{figure}

We show in Fig.\,\ref{fig:d68-g2} the behaviour of $d_6$ and $d_8$ versus $\la\alpha_s G^2\ra$. One can notice that:

\d The value of the four-quark condensate estimated from factorization is inconsistent with the SVZ value of the gluon condensate.

\d  The ratio:
\beq
r_{46} \equiv \frac{\rho\la\bar\psi\psi\ra^2}{\la\alpha_s G^2\ra}\approx 0.9\times 10^{-2}~{\rm GeV^2},
\eeq
is almost constant.  The value of  $r_{46} $ obtained here  has also been obtained from different approaches\,\cite{LNT,FESR,SOLA,BORDES,CAUSSE} independently on the absolute size of the extracted condensates and on the kind of methods.  However, we notice that  the $\tau$-like moments values given in Table\,\ref{tab:g2} do not fulfill such relations. 

\section{Confronting theory and experiment for ${\cal R}^{ee}_{10}$ and ${\cal L}^{ee}_0$}
Using as input the previous values of the condensates and the value of $\alpha_s$ from PDG, we compare the QCD and experimental sides of  ${\cal R}_{10}$ and ${\cal L}_0$ which we show in Figs.\,\ref{fig:Ree} and \ref{fig:Lee}.
\subsection*{\b Ratio of moments ${\cal R}^{ee}_{10}$}
\begin{center}
\begin{figure}[hbt]
\begin{center}
\includegraphics[width=15.cm]{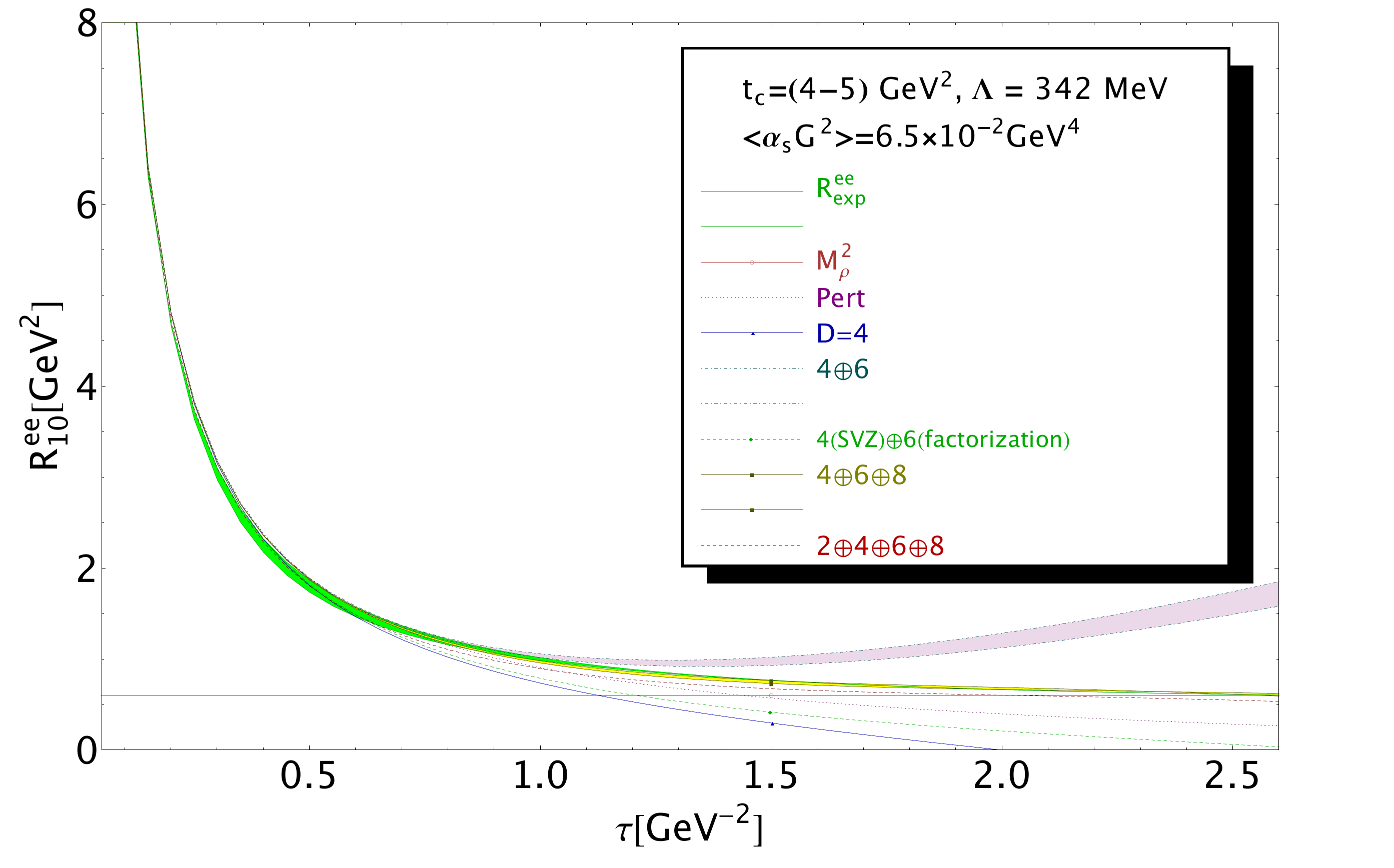}
\caption{\footnotesize  $\tau$-behaviour of the ratio of moments ${\cal R}^{ee}_{10}$ for different truncation of the OPE.}\label{fig:Ree}.
\end{center}
\vspace*{-0.5cm}
\end{figure} 
\end{center}

We show the QCD predictions for different truncation of the OPE in Fig.\,\ref{fig:Ree}. We observe that:

\d For $\tau\leq 1.5$ GeV$^{-2}$, the truncation of the OPE up to $D=6$ gives a good description  of the data where we have a minimum. The presence of minimum or inflexion point is expected for an approximate QCD series as discussed in the example of an harmonic oscillator\,\cite{BELLa,BERTa,SNB1,SNB2}. The use of the standard SVZ value of the gluon condensate $\oplus$ the value of the four-quark condensate estimated using the factorization assumption underestimates the data. However, the simultaneous use of the SVZ standard value $\oplus$ the factorization (combination often used in the QCD sum rules literature) is inconsistent from the result of simultaneous two-parameter fit   shown in the previous section and in Fig.\,\ref{fig:d68-g2}.

\d The inclusion of the $D=8$ condensate enlarges the region of agreement between the QCD prediction and the experiment until $\tau=2.5$ GeV$^{-2}$ where the $\rho$-meson mass is reached and where, exceptionally, the PT series for the spectral function still make sense. 

\d We complete the analysis by adding the tachyonic gluon mass beyond the SVZ-expansion. We see that it tends to decrease the agreement with the data but the effect is (almost) negligible within our precision. 

\subsection*{\b Lowest moment ${\cal L}^{ee}_{0}$}
This moment has been used in Refs.\,\cite{EIDELMAN,SN95} to estimate $\alpha_s$ and the QCD condensates. 

\hspace*{0.5cm} \d We have tried to extract these parameters using ${\cal L}^{ee}_{0}$ but the results are unconclusive. Contrary to the case of 
${\cal R}^{ee}_{10}$, we notice an important sensitivity of the results on the truncation of the PT series. This problem can be evaded if one uses the $\pi^2$ resummation of the higher order term where the PT series converges faster\,\cite{KAHN}. However, we also remark that the determination from the low and high sets of data vary in a large range leading to a very inaccurate result. 

\begin{center}
\begin{figure}[hbt]
\begin{center}
\includegraphics[width=15.cm]{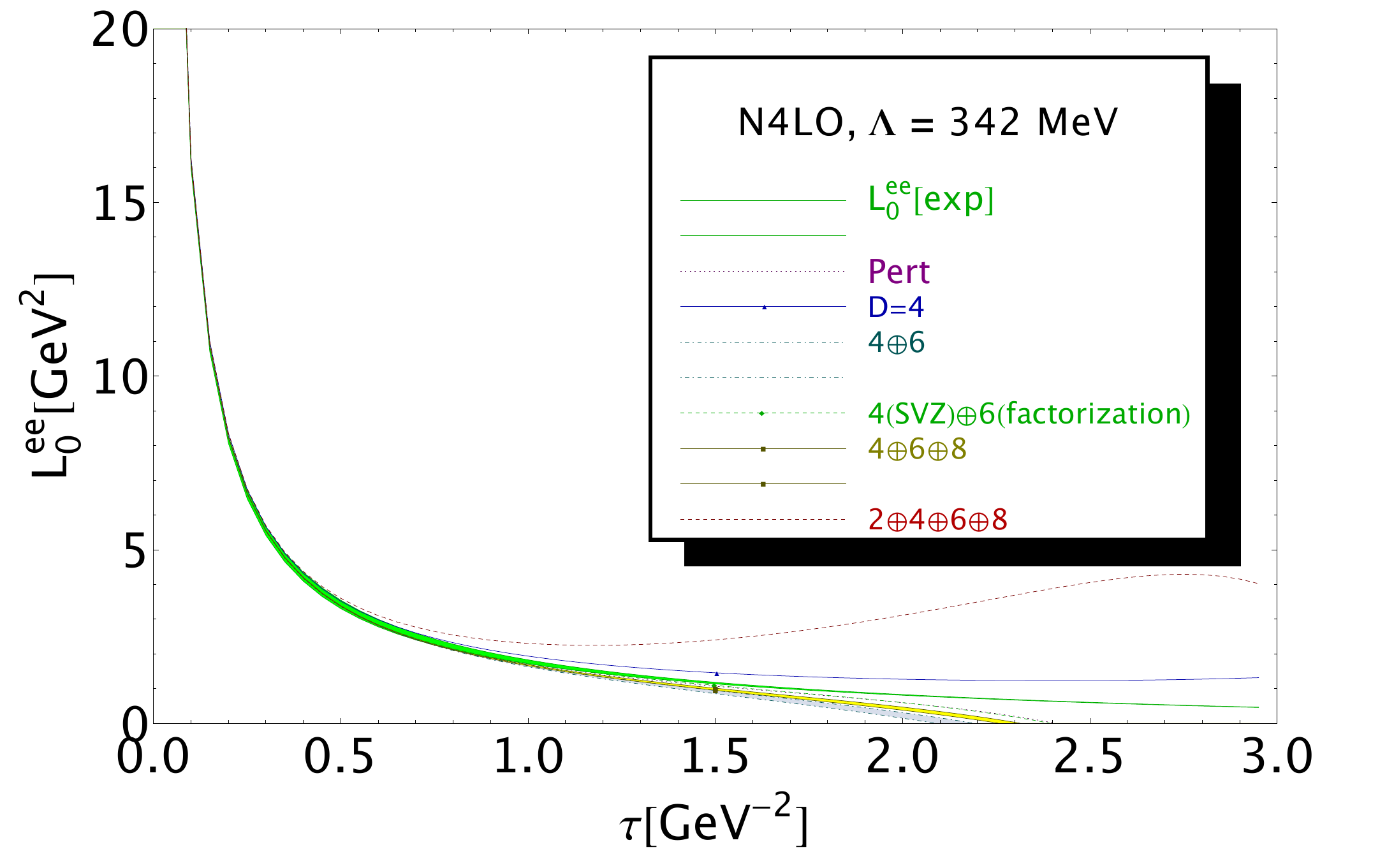}
\caption{\footnotesize  $\tau$-behaviour of the lowest moment ${\cal L}_0$ for different truncation of the OPE.}\label{fig:Lee}
\end{center}
\vspace*{-0.5cm}
\end{figure} 
\end{center}

\hspace*{0.5cm} \d Therefore, we just use the previous values of $\alpha_s$ from the PDG average\,\cite{PDG}, $\la\alpha_s G^2\ra$ from Eq.\,\ref{eq:asg2} and $d_4$ and $d_6$ obtained in the previous section and compare the experimental and theoretical $\tau$-behaviour of the  lowest moment ${\cal L}^{ee}_{0}$ in Fig.\,\ref{fig:Lee}. 

\hspace*{0.5cm} \d We observe that  the effect of the:

--  Tachyonic gluon mass, if any,  is quite large for $\tau\geq 0.6$ GeV$^{-2}$. 

--  $D=8$ condensate is relatively small due to the (1/6) factor suppression induced by the exponential weight for the Laplace / Borel sum rule. 

-- Combination  of the standard value of the SVZ condensate and of factorization for the four-quark condensate gives a better agreement with the data here than in the case of the ratio of moments. However, as mentioned previously, a simultaneous use of these two values is (unfortunately)  inconsistent with the fitted ones from different approaches. 

  \section{$\alpha_s$ from the lowest $\tau$-like decay moment}
  $\tau$-like decay moments applied to $e^+e^-\to$ Hadrons have been initially used in Refs.\,\cite{SNPICH,SN95} to extract $\alpha_s$ and the QCD condensates. 
  Using the previous values of the QCD non-perturbative (NP) parameters into the $\tau$-like decay rate (lowest moment)\,\cite{BNP}:
  \beq
  R^{ee}_{\tau}=\int_0^{1} dx_0\, (1-3x_0^2+2x_0^3)2R^{I=1}_{ee}(x_0)
  \eeq
  with $x_0\equiv (t/M^2_0)$, we attempt to extract the value of $\alpha_s$ for different values of $M_0$.
  \subsection*{\b The QCD expression of $ R^{ee}_{\tau}$}
 We shall use its expression  within Fixed Order perturbation theory (FO)\,\footnote{There are some recent attempts to reconcile FO and Contour Improved (CI) approach (see e.g. \,\cite{JAMIN}).}. In our analysis, we shall omit non-standard effects due to  instantons and duality violation where the effects are expected to be negligible\,\cite{PICH2,SNTAU}.  Renormalon ones induced by the large $\beta$ approximation of the higher order terms of PT\,\cite{RENORM,AYALA} will not also be considered like its tachyonic gluon mass effects\,\cite{ZAKA} and Pad\'e approximants alternative. However, their eventual effects are expected to be smaller than the size of the errors in our analysis. 
  
  Following the notations of BNP, the QCD expression of the moment can be written as:
  \beq
  R^{ee}_{\tau}\vert_{qcd}= \frac{N_c}{2}\Big{[} 1+\delta_{ee}^{0}+\sum_{D=1,2,\dots}\delta_{ee}^{(2D)}\Big{]}.
  \eeq
 \d The QCD expression copied from BNP\,\cite{BNP} reads within FO perturbation theory in the $\overline{MS}$ scheme\,:
  \beq
  \delta^{(0)}_{ee}\vert_{FO}=a_s+5.2023\,a_s^2+26.366\,a_s^3+ 127.079\,a_s^4,
  \eeq
  where the last term comes from\,\cite{CHET4}.  The power corrections read:
  \bea
   \delta^{(2)}_{ee}&=&-12(1+4a_s)\frac{(m_u^2+m_d^2)}{M_0^2},\,\,\,\,\,\,\,\,\,\,\,\,\,\,\,\,\,\,\,\,\,\,\,\,\,\,\,\,\,\,\,\,\,\,\,\,\,
   \delta^{(2)}_{ee}\vert_{tach}=-2\times 1.05\frac{\,a_s\lambda^2}{M_0^2},\nnb\\
   \delta^{(4)}_{ee}&=&\frac{11\pi}{4}a_s^2\frac{\la\alpha_s G^2\ra}{M_0^4}
  -\frac{2\pi^2a_s^2}{M_0^4}\Big{[} 9\la m_u\bar \psi_u\psi_u++m_d\bar\psi_d\psi_d \ra+4\sum_k \la m_k\bar \psi_k\psi_k\ra\Big{]}+36\frac{(m_u^4+m_d^4)}{M_0^4}\Big{]},\nnb\\
   \delta^{(6)}_{ee}&=& -6\frac{d_6}{M_0^6},\,\,\,\,\,\,\,\,\,\,\,\,\,\,\,\,\,\,\,\,\,\,\,\,\,\,\,\,\,\,\,\,\,\,\,\,\,\,\,\,\,\,\,\,\,\,\,\,\,\,\,\,\,\,\,\,\,\,\,\,\,\,\,\,\,\,\,\,\,\,\,\,\,\,\,\,\,\,\,\,\,\,\,
   \delta_{ee}^{(8)}= -4\frac{d_8}{M_0^8},
  \eea
  where $d_{6}$ and $d_{8}$ have been defined in Eqs.\,\ref{eq:d6} and \ref{eq:d8}. We have shown the $D=2$ contribution due to an eventual tachyonic gluon mass not included in the standard OPE. 
  
  \d Using the Contour Improved (CI) approach, the QCD PT series read\,\cite{LEDI,CHET4}:
    \beq
  \delta^{(0)}_{ee}\vert_{CI}=1.364\,a_s+2.54\,a_s^2+9.71\,a_s^3+ 64.29\,a_s^4,
  \eeq
  \d  Assuming that the PT series grow geometrically\,\cite{SZ} as observed from the calculated coefficients, we estimate the $a_s^5$ coefficient to be :
  \beq
 p_5^ {FO}\approx +597, ~~~~~~~ ~~~~~~~~~~~~~~~p_5^{CI}\approx +426~,
  \label{eq:as5}
  \eeq
 which we consider either to be  the error due to the higher order terms of the series or (more optimistically) to be the estimate of the uncalculated $\alpha_s^5$ coefficient.  A check of the sign and size of this estimate  from another approach is welcome to test if the PT series already reach its asymptotic at this order and if the large $\beta$-approach  is a good approximation. 
  \subsection*{\b Numerical analysis}
 \d  We insert into   $R^{ee}_{\tau}\vert_{qcd}$ the value of the condensates given in Eqs.\,\ref{eq:asg2} and \ref{eq:res-d68}. We extract $\alpha_s$ for different values of $M_0$ using the Mathematica Program NSolve. Next we run each result to $M_\tau$ using the evolution equation:
  \beq
  a_s(M_\tau)=a_s(M_0)\Bigg{\{}1+ a_s(M_0)\,\log (x_0)\Bigg{[} \frac{\beta_1}{2}+a_s(M_0)\ga \frac{\beta_2}{2}
  +\frac{\beta_1^2}{4}\log (x_0)\dr \Bigg{]}\Bigg{\}}+{\cal O}(a_s^4)
  \eeq
  
\begin{center}
\begin{figure}[hbt]
\begin{center}
\includegraphics[width=11.cm]{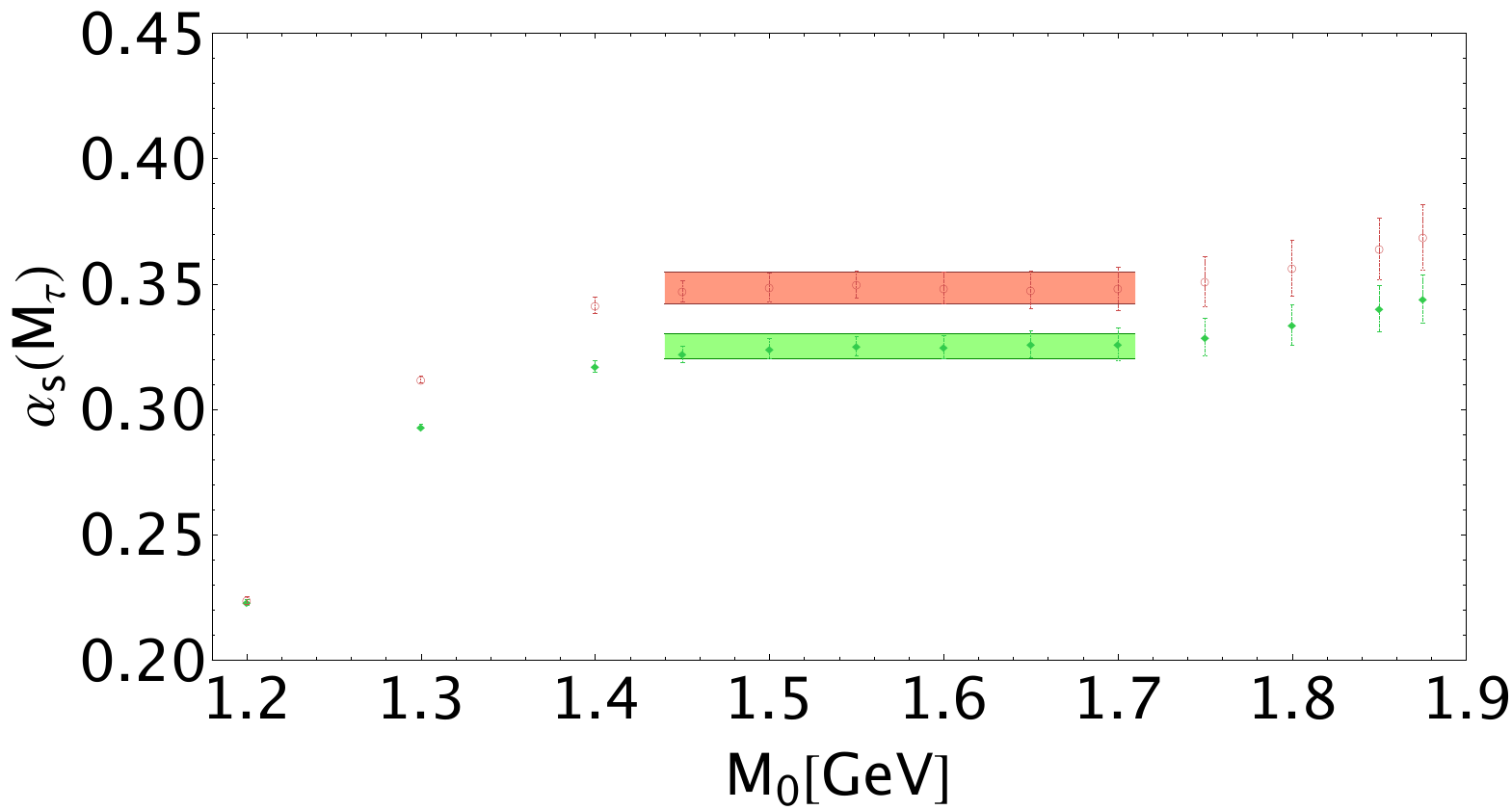}
\caption{\footnotesize Value of $\alpha_s(M_\tau)$ to ${\cal O}(\alpha_s^4)$ as a function of the hypothetical $\tau$-mass $M_0$ including $D\leq 8$ condensates. Green : FO. Red : CI. The errors due to the non-perturbative condensates and to the fitting procedure are shown.}\label{fig:as}.
\end{center}
\vspace*{-0.5cm}
\end{figure} 
\end{center}

  \d We show the different values of $a_s(M_\tau)$ extracted at each value of $M_0$ in Fig.\,\ref{fig:as}. We notice a nice stability of the results versus $M_0$ in the range 1.45 to 1.7 GeV. We obtain to ${\cal O}(\alpha_s^4)$\,:
  \bea
  \alpha_s^{FO}(M_\tau)&=&0.3247 (46)_{fit}(1)_{np} (62)_{h.o}=0.3247 (77)~~~\to~~  \alpha_s^{FO}(M_Z)=0.1191(10)(3)\, \nnb\\
   \alpha_s^{CI}(M_\tau)&=&0.3483 (63)_{fit}(1)_{np} (15)_{h.o}=0.3483 (65)~~~\to~~  \alpha_s^{CI}(M_Z)=0.1218(8)(3)\, ,
  \label{eq:alphas}
  \eea
 where the dominant error comes from the fitting procedure and the estimate of the higher order corrections.  The last error in $\alpha_s(M_Z)$ comes from the running procedure. We deduce the mean\,:
 \beq
  \alpha_s(M_\tau)=0.3385 (50)(136)_{syst}~~~\to~~  \alpha_s(M_Z)=0.1207(17)(3)\, ,
  \eeq
  where $\pm 136$ is an added conservative systematic from the distance of the mean to the FO/CI central values. $\pm 3 $ is
  an error induced by the running procedure. 
  
  \d  The error due to the non-perturbative (n.p) terms are  negligible for FO and CI. 
  
\d We have not included the effects beyond the standard OPE (tachyonic gluon, instantons, duality violation) in our analysis which have been estimated to be small\,\cite{SN95}. The previous values of $\alpha_s(M_\tau)$ are in agreement, within the errors, with some other determinations from $V/A/V+A$ hadronic $\tau$-decay data for FO  and CI within the standard OPE (see  Table\,\ref{tab:alpha_s}). 

\d We have not included in Table\,\ref{tab:alpha_s} some analysis including a duality violation  \,\cite{BOITO,PICH2} and PT $\oplus$ renormalon\,\cite{AYALA,RENORM} which  leads to smaller values of $\alpha_s(M_\tau)$.  The mean of different FO/CI determinations up to ${\cal O}(\alpha_s^4)$ in Table\,\ref{tab:alpha_s} is\,:
\beq
 \la \alpha_s(M_\tau)\ra =0.3336 (31)(116)_{syst}~~~\to~~\la  \alpha_s(M_Z)\ra=0.1201(14)(3)\, .
  \eeq

  \d We notice that the size of the previously estimated higher order term (h.o) for FO is about the same as coming from renormalon or Pad\'e approximants\,\cite{RENORM,AYALA}. Taken literally, the estimated $\alpha_s^5$ coefficients in Eq.\,\ref{eq:as5},  the value of $\alpha_s$ becomes to ${\cal O}(\alpha_s^5)$\,:
  \bea
   \alpha_s^{FO}(M_\tau)\vert_{+h.o}&=&0.3185 (46)~~~\to~~  \alpha_s^{FO}(M_Z)=0.1184(9)\, \nnb\\
   \alpha_s^{CI}(M_\tau)\vert_{+h.o}&=&0.3401(61)~~~\to~~  \alpha_s^{CI}(M_Z)=0.1209(7)\,,
\eea
which gives the mean\,:
\beq
   \alpha_s(M_\tau)= 0.3263(37)(78)_{syst}=0.3263(86)~~~\to~~  \alpha_s(M_Z)=0.1193(11)(3)\, 
  \eeq

   {\footnotesize
\begin{table}[hbt]
\setlength{\tabcolsep}{1.1pc}
  \begin{center}
    {
  \begin{tabular}{lllll}

&\\
\hline
\hline
$\alpha_s ^{FO}(M_\tau)$&$\alpha_s ^{CI}(M_\tau)$&$\delta^V_{NP}(M_\tau)\times 10^{-2}$  &Data& Refs.\\
 \hline 
\boldmath$0.3247(77)$&\boldmath$0.3483(65)$&\boldmath$2.30\pm 0.20$&\boldmath$e^+e^-$&{\bf This work}\\
0.350(50)&--&$1.45\pm 1.3$ &--& \cite{SNPICH} \\
0.320(30)&--& $3.60\pm 1.64$&  $e^+e^-\oplus$ V+A\,: $\tau$-decay & \cite{SN95}\\
0.320(22)&0.340(23)&$2.0\pm1.7$&V\,: $\tau$-decay&\,{\rm ALEPH}\,\cite{ALEPH}\\
0.323(16)&0.347(23)&$1.87\pm 0.54$&V\,: $\tau$-decay&{\rm OPAL}\,\cite{OPAL}\\
0.328(9)&--&--& V+A\,:  $\tau$-decay & \cite{SNTAU}\\
0.322(16)&0.342(16)&-- &V+A\,: $\tau$-decay&\cite{CHET4}\\
0.324(11)$^{*)}$&0.346(11) &--&V+A\,: $\tau$-decay&\,{\rm ALEPH}\,\cite{DAVIER}\\
0.320(12)&0.335(13)&--&V+A\,: $\tau$-decay&\cite{PICH2}\\
\hline
\it  0.3243(42) &\it 0.3452(47)&$\it 2.28\pm 0.20$&& \it Mean \\
   \hline\hline
\end{tabular}}
 \caption{ ~$\alpha_s(M_\tau)$ and $\delta^V_{NP}$ from $\tau$-decay moments within fixed order (FO) and contour improved (CI) perturbative series up to ${\cal O}(\alpha_s^4)$. $^{*)}$ indicates that the quoted error is our crude estimate. }\label{tab:alpha_s} 
 \end{center}
\end{table}
} 

  \section{Power corrections to the lowest $\tau$-decay moment $R^{ee}_{\tau}$}
  
\subsection*{\b $d_6$  from a one-parameter fit}
\begin{figure}[hbt]
\begin{center}
\includegraphics[width=9.cm]{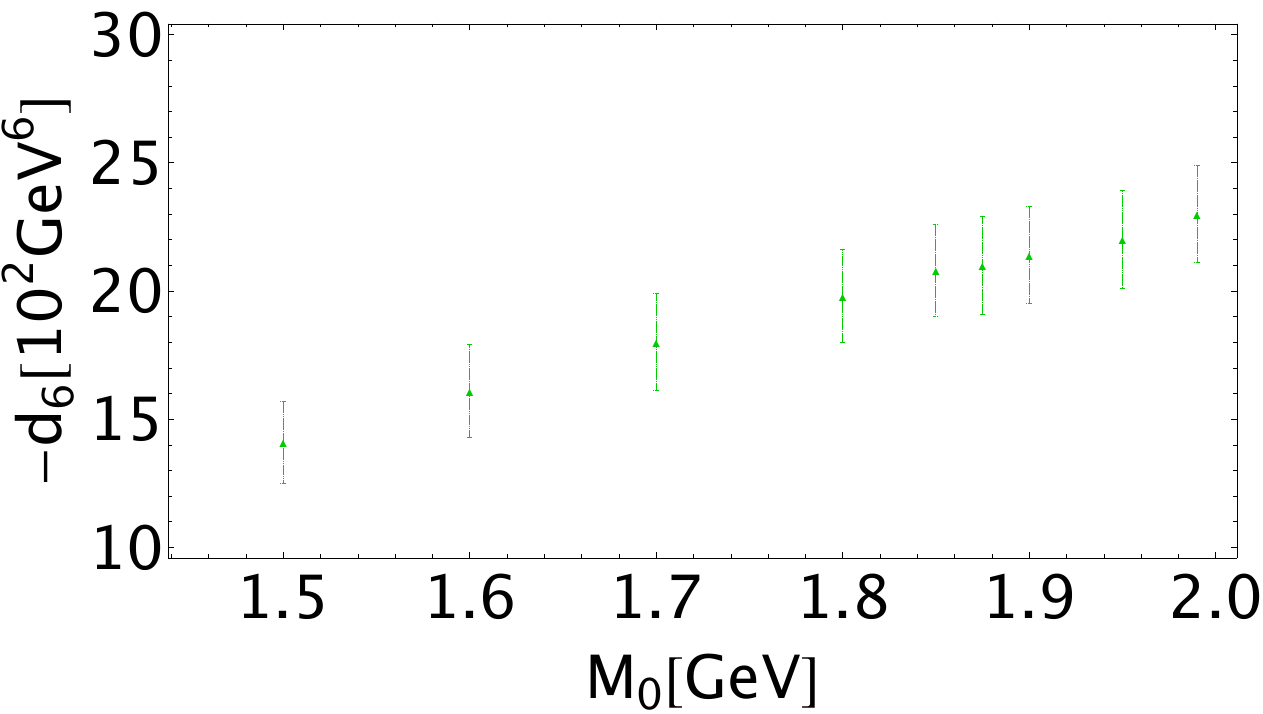}
\caption{\footnotesize $d_6$ condensate as function of the hypothetical $\tau$-mass $M_0$ using $\alpha_s(M_0), d_4$ and $d_8$ from ${\cal R}_{10}$ as inputs.}\label{fig:d6-tau}.
\end{center}
\vspace*{-0.5cm}
\end{figure}

To give a stronger constraint on $d_6$, we shall work with a one-parameter fit.   We shall use as input the value of the gluon condensate in Eq.\,\ref{eq:asg2} and the previously determined values of $\alpha_s(M_\tau)$ from $R^{ee}_{\tau}$ in Eq.\,\ref{eq:alphas} and  of $d_8$ in Eq.\ref{eq:res-d68}  from ${\cal R}_{10}$. 
The result of the analysis is shown in Fig.\,\ref{fig:d6-tau} where one remarks that the result increases softly with the value of $M_0$ and shows  an almost stability (inflection point) around (1.8-1.9) GeV which is: 
\beq
d_6\simeq -(20\sim 21)\times 10^{-2}~{\rm GeV}^6
\eeq
with a 10\% error. This result is identical to the one in Eq.\,\ref{eq:res-d68}. However, the results obtained from some other $\tau$-decay moments are also recovered in the $M_0$-instability region.  We conclude that the extraction of $d_6$ from the low $\tau$-decay moment is less accurate than the one from $R_{10}^{ee}$. It only gives an approximate range of values. 


  \subsection*{\b Sum of power corrections to the lowest $\tau$-decay moment $R^{ee}_{\tau}$}

\d  Using our previous determinations of the condensates, we can estimate the sum of the power corrections to $R^{ee}_{\tau}$.
   In so doing, we introduce as input the value of $\alpha_s(M_\tau)$ determined previously and the value of the gluon condensate in Eq.\,\ref{eq:asg2}. To minimize the number of free parameters, we first neglect the contribution of high-dimension condensates $D\geq 10$ and assume that the retained condensate $D=8$ is an effective condensate which absorbs into it all unknown high-dimension condensate effects. 
   
 
\d  Then, using the values of $d_4$ in Eq.\ref{eq:asg2}, $d_6$ and $d_8$ in Eq.\,\ref{eq:res-d68}, we deduce the sum of the NP contributions\,:
\beq
\delta^V_{NP}(M_\tau)= (2.3\pm 0.2)\times 10^{-2},
\eeq
which improves our previous findings from $e^+e^-\to I=1$ hadrons data\,: $\delta^V_{NP}(M_\tau)= (2.38\pm 0.89)\times 10^{-2}$ in Ref.\,\cite{SN95} and the analysis in\,\cite{SNTAU}.  

\d It is remarkable to notice that this value agrees with the one from $\tau$-decay analysis (see Table\,\ref{tab:alpha_s}) despite the large discrepancies with the individual values of each condensate.  This is due to the alternate signes of the condensate contributions in the $\tau$-moments.

\d One can also notice that the value of $\alpha_s$ decreases with the $\delta_{NP}$.  From $\delta^V_{NP}=3.7\times 10^{-2}$  (our first iteration in Eq.\,\ref{eq:d68-min}) to the final value: $2.3 \times 10^{-2}$ (Eqs.\,\ref{eq:res-d68} and \ref{eq:d8-res}), the value of $\alpha_s(M_\tau)$ moves slightly  from 0.329 to 0.325. 

\section{$a_\mu\vert^{hvp}_{l.o}$ from   $e^+e^-\to$ hadrons data}
We complete the analysis by updating our previous determination of the  lowest order hadronic contributions  to the vacuum polarization of $a_\mu$\,\cite{SNamu}.  This analysis will also serve as a test of our parametrization of the $I=1$ part of the spectral function used in previous sections. 

The lowest order hadronic contributions to the vacuum polarization can be obtained from the
well-known dispersion relation\,\cite{BOUCHIAT,DURAND,KINOSHITA,BOWCOCK,GOURDIN}\,:
\beq
a_\mu\equiv \frac{1}{2}(g-2)_\mu= \frac{1}{4\pi^3}\int_{4m_\pi^2}^\infty dt\, K_\mu(t)\sigma (e^+e^-\to {\rm hadrons})
\eeq
where $K_\mu(t)$ is the QED kernel function\,\cite{LAUTRUP}:
\beq
K_\mu(t)=\int_0^1dx\frac{x^2(1-x)}{x^2+(t/m_\mu^2)(1-x)}.
\label{eq:kmu}
\eeq
\subsection*{\b $a_\mu\vert^{hvp}_{l.o}$ from   $e^+e^-\to I=1$ hadrons below 1.875 GeV and test of our parametrization}
This analysis is an update of the one in Ref.\,\cite{SNamu} and will also serve as a test of our parametrization of the data especially for the $\rho$-meson where we have only used a minimal Breit-Wigner for the pion form factor. We obtain the contributions given in Table 3 for the three subdivision regions of the data below 1.875 GeV as shown in Figs.\,\ref{fig:rho}, \ref{fig:3pi} and \ref{fig:rhoprime}.

\begin{figure}[hbt]
\begin{center}
\includegraphics[width=11.cm]{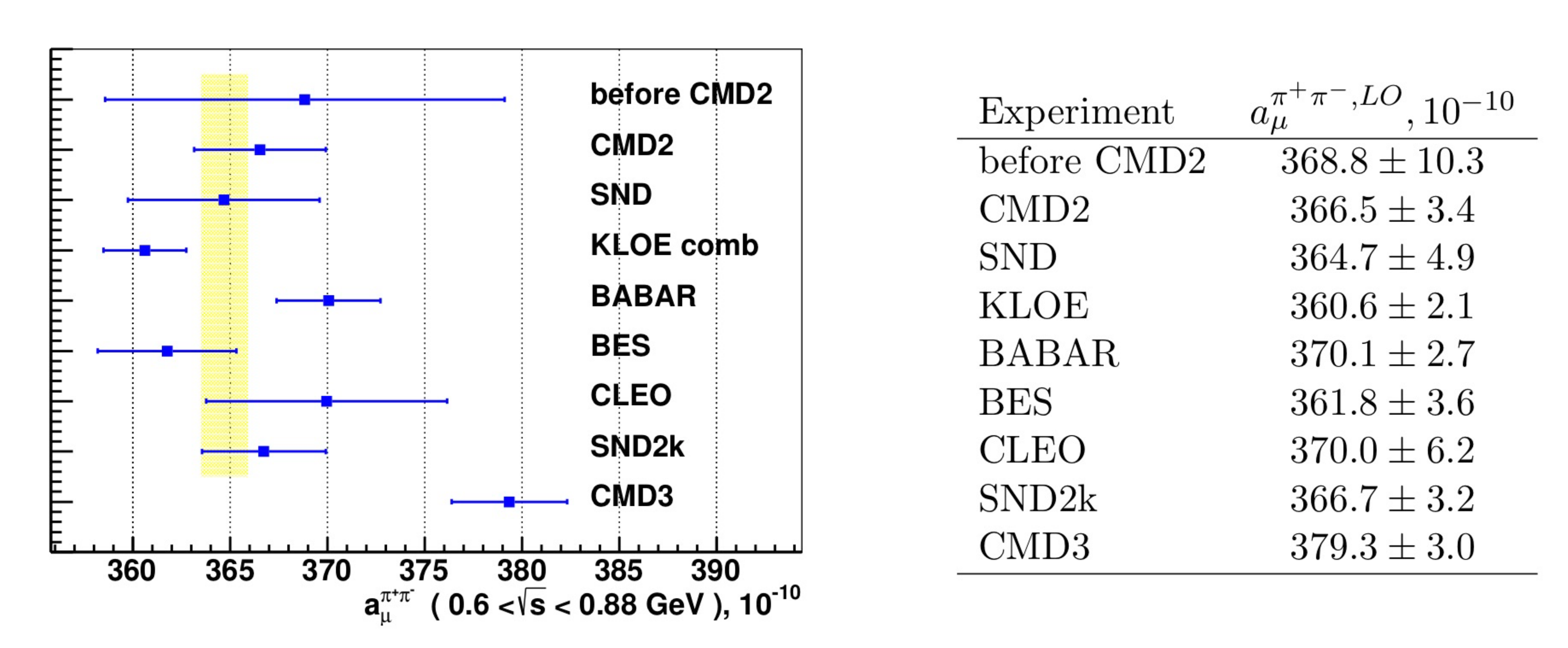}
\vspace*{-0.5cm}
\caption{\footnotesize  Comparison of the $\rho$-meson contribution to $a_\mu$ for $0.6\leq \sqrt{s}\leq 0.88$ GeV from \cite{CMD3}. } \label{fig:amu}
\end{center}
\vspace*{-0.75cm}
\end{figure} 

   {\footnotesize
\begin{table}[H]
\vspace*{-1.cm} 
\setlength{\tabcolsep}{1.pc}
  \begin{center}
    { \footnotesize
  \begin{tabular}{| llll |}
\hline
\hline
\rowcolor{yellow}\boldmath$\sqrt{t}$\,\bf [GeV] &\boldmath$a_\mu\vert^{hvp}_{l.o}\times 10^{11} $&\boldmath$a_\tau\vert^{hvp}_{l.o}\times 10^{9} $&$\Delta\alpha^{(5)}_{had}(M_Z^2)\times 10^5$\\ 
 \hline 

{\bf Light I=1}&&&\\
$\rho(2m_\pi\to 0.50)$&$489.4\pm 3.1$&$85.1\pm 0.6$&$12.38\pm 0.08$\\
$\rho(0.50\to 0.60)$&$524.5\pm 13.7$&$135.9\pm 3.5$&$22.84\pm 0.59$\\
$\rho(0.60\to 0.776)$&$2712.0\pm 30.2$&$943.7\pm10.3$&$182.87\pm 1.97$ \\

$\rho(0.776\to 0.993)$&$1297.3\pm9.8$&$1797.3\pm 20.3$&$117.82\pm 0.98$\\
$0.993\to 1.5$&$354.4\pm 6.7$&$228.2\pm 4.1$&$67.3\pm 1.14$\\
$1.5\to 1.875$&$237.6\pm 5.7$&$206.8\pm 4.9$&$80.11\pm 1.9$\\

{\it Total Light I=1 ($\leq 1.875$)}&$\it 5615.2\pm 36.0$&$\it 2453.3\pm 11.9$&$\it 483.3\pm 4.4$\\
{\bf Light I=0}&&&\\
$\omega$ (NWA) &$417.1\pm 13.7$&$163.3\pm 5.3$&$33.6\pm 1.1$\\
$\phi$ (NWA) &$389.6\pm 4.6$&$20.6\pm 0.2$&$51.2\pm 0.6$\\
$0.993\to 1.5$ & $44.3\pm 0.8$&$28.5\pm 0.5$&$8.4\pm 0.1$\\
$\omega(1650)$(BW)&$24.3\pm 0.1$&$16.7\pm 0.1$&$5.2\pm 0.1$ \\
$\phi(1680)$(BW)&$1.8\pm 0.9$&$1.3\pm 0.6$&$0.4\pm 0.2$ \\
{\it Total Light I=0 ($\leq 1.875$)}&$\it 877.1\pm 14.5$&$\it 230.4\pm 5.3$&$\it 99.4\pm 1.3$\\
{\bf Light I=\boldmath $0\, \oplus \,1$}&&&\\
$1.875\to 1.913$ & $14.7\pm 0.7$&$14.4\pm 0.7$&$6.3\pm 0.3$\\
$1.913\to 1.96$ & $17.6\pm 0.6$&$17.5\pm 0.6$&$7.9\pm 0.3$\\
$1.96\to 2$ & $14.3\pm 0.5$&$14.5\pm 0.5$&$6.7\pm 0.2$\\
$2\to 3.68$:  QCD $(u,d,s)$ & $247.2\pm 0.3$&$308.3\pm 0.5$&$202.8\pm 0.5$\\
{\it Total Light I=$0\, \oplus \,1$ ($1.875\to 3.68$)}&$\it 293.8\pm 1.1$&$\it 354.7\pm 1.2$&$\it 223.7\pm 0.7$\\
\rowcolor{greenli}{ Total Light I=$0\, \oplus \,1$ ($2m_\pi\to3.68$)}&$ 6786.1\pm 38.8$&$3038.4\pm 24.5$&$806.4\pm 4.6$\\
{\bf Charmonium} &&&\\
$J/\psi (1S)$ (NWA) &$65.1\pm 1.2$&$92.7\pm 1.8$&$73.5\pm 1.4$\\
$\psi (2S)$ (NWA) &$16.4\pm 0.6$&$26.0\pm 0.9$&$26.1\pm 0.8$\\
$\psi (3773)$ (NWA) &$1.7\pm 0.1$&$2.7\pm 0.2$&$2.9\pm 0.2$\\
{\it Total $J/\psi (NWA)$} &$\it 83.2\pm 1.4$&$\it 121.4\pm 2.0$&$\it 102.5\pm 1.6$\\
$3.69\to 3.86$ & $11.4\pm 1.0$&$18.3\pm 1.6$&$19.0\pm 1.6$\\
$3.86\to4.094$ & $16.6\pm 0.5$&$27.5\pm 0.8$&$30.9\pm 0.9$\\
$4.094\to4.18$ & $6.6\pm 0.2$&$11.2\pm 0.3$&$13.2\pm 0.4$\\
$4.18\to4.292$ & $6.5\pm 0.2$&$11.2\pm 0.4$&$13.7\pm 0.5$\\
$4.292\to4.54$ & $11.8\pm 0.6$&$20.7\pm 0.8$&$26.8\pm 1.1$\\
$4.54\to10.50$:  QCD $(u,d,s,c)$ & $92.0\pm 0.0$&$186.2\pm 0.0$&$458.7.3\pm 0.1$\\
{\it Total Charmonium ($3.69\to 10.50$)}&$\it 145.1\pm 1.3$&$\it 275.6\pm 2.0$&$\it 564.9\pm 2.2$\\
\rowcolor{greenli}{ Total Charmonium }&$ 228.1\pm 1.9$&$396.5\pm 2.8$&$664.8\pm 2.7$\\
 {\bf Bottomium} &&&\\
$\Upsilon (1S)$ (NWA) &$0.54\pm 0.02$&$1.25\pm 0.07$&$5.65\pm 0.29$\\
$\Upsilon(2S)$ (NWA) &$0.22\pm 0.02$&$0.51\pm 0.06$&$2.54\pm 0.29$\\
$\Upsilon (3S)$ (NWA) &$0.14\pm 0.02$&$0.33\pm 0.04$&$1.77\pm 0.23$\\
$\Upsilon(4S)$ (NWA) &$0.10\pm 0.01$&$0.23\pm 0.03$&$1.26\pm 0.16$\\
$\Upsilon(10.86\,\oplus\,11)$ (NWA) &$0.1\pm 0.06$&$0.20\pm 0.06$&$1.67\pm 0.39$\\
{\it Total Bottomium (NWA)}&$\it1.0\pm 0.1$&$\it 2.3\pm 0.1$&$\it 11.2\pm 0.5$ \\
$Z-pole $&-&-&29.2\,\cite{YND17}\\
$10.59\to2m_t$\,:  QCD $(u,d,s,c,b)$ & $22.4\pm 0.3$&$57.5\pm 0.1$&$1282.9\pm 1.2$\\
\rowcolor{greenli}{Total Bottomium }&$ 23.4\pm 0.3$&$59.8\pm 0.1$&$1323.3\pm1.3$\\
$2m_t\to\infty$\,: QCD $(u,d,s,c,b,t)$&0.03&0.08&-28.2\\
\hline
\rowcolor{yellow}\bf Total sum &\boldmath$7036.5\pm 38.9$&\boldmath$3494.8\pm 24.7$&\boldmath$2766.3\pm 4.5$\\

   \hline\hline
 
\end{tabular}}
 \caption{ 
  $a_\tau\vert^{hvp}_{l.o}$, $a_\mu\vert^{hvp}_{l.o}$ and $\alpha(M^2_z)$ within our parametrization of the compiled PDG\,\cite{PDG} $\oplus$ the recent CMD3\,\cite{CMD3}. }\label{tab:amu1} 
 \end{center}
\end{table}
} 
\subsection*{\hspace*{0.5cm}\d Comparison with some other determinations for $0.60\leq \sqrt{t}\leq 0.88$ GeV}
We compare our determinations from the $\rho$-meson using a minimal Breit-Wigner parametrization of the pion form factor for $0.6\leq \sqrt{t}\leq 0.88$ GeV\,:
\beq
a_\mu^{\rho}\vert^{hvp}_{l.o}[0.6\to 0.88]=(370.5\pm 2.1)\times 10^{-10}.
\label{eq:compare}
\eeq
\vspace*{-0.10cm}
 Despite our minimal parametrization of the pion form factor, our result is in a good agreement 
 with the previous ones in Fig.\ref{fig:amu} using  a more involved parametrization. We interpret our slightly lower value compared to CMD3\,\cite{CMD3} from the fact that the minimal fit cannot accommodate accurately the data in this region.  

\subsection*{\hspace*{0.5cm} \d  Improving the $\rho$ meson contribution below 0.993 GeV}
\begin{figure}[hbt]
\begin{center}
\includegraphics[width=8.cm]{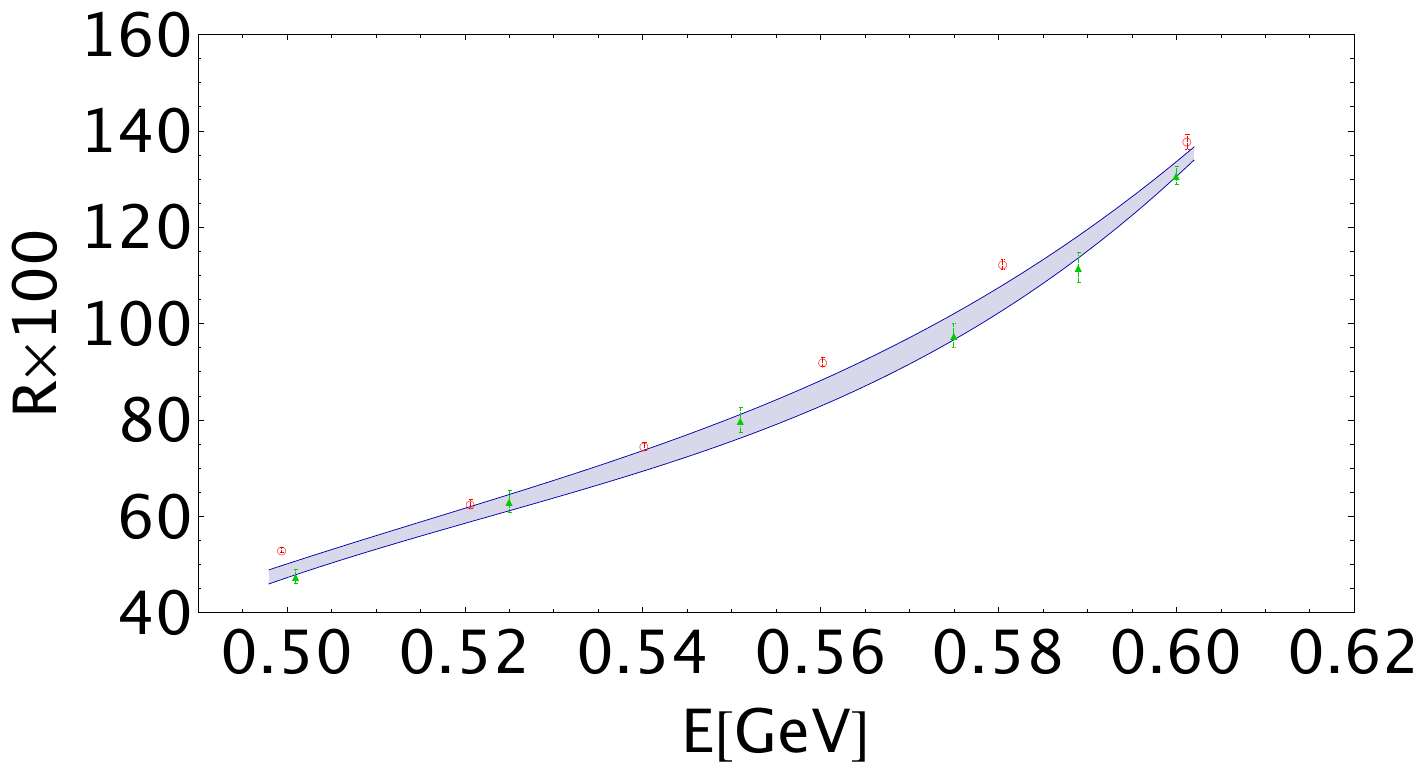}
\includegraphics[width=8.cm]{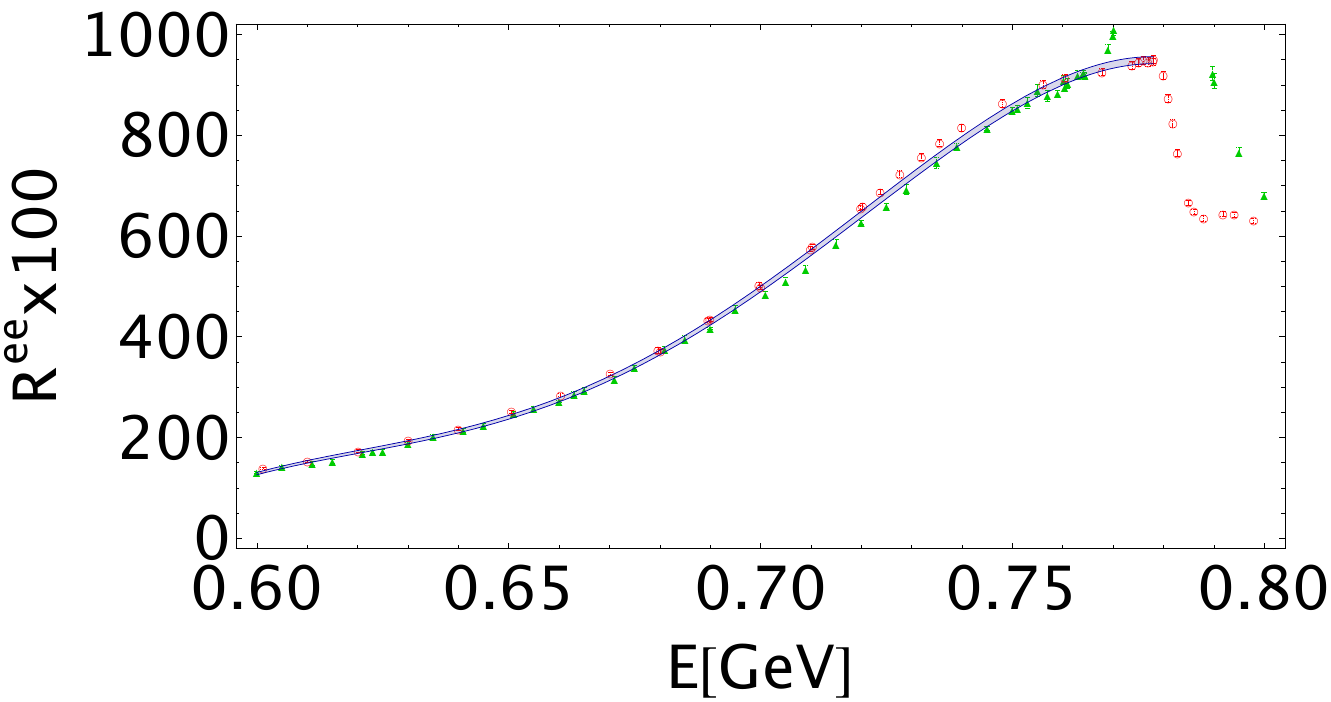}
\caption{\footnotesize  Fit of the data in the region 0.5 to 0.776 GeV. } \label{fig:rho2}
\end{center}
\vspace*{-0.5cm}
\end{figure}

\begin{figure}[hbt]
\begin{center}
\includegraphics[width=8cm]{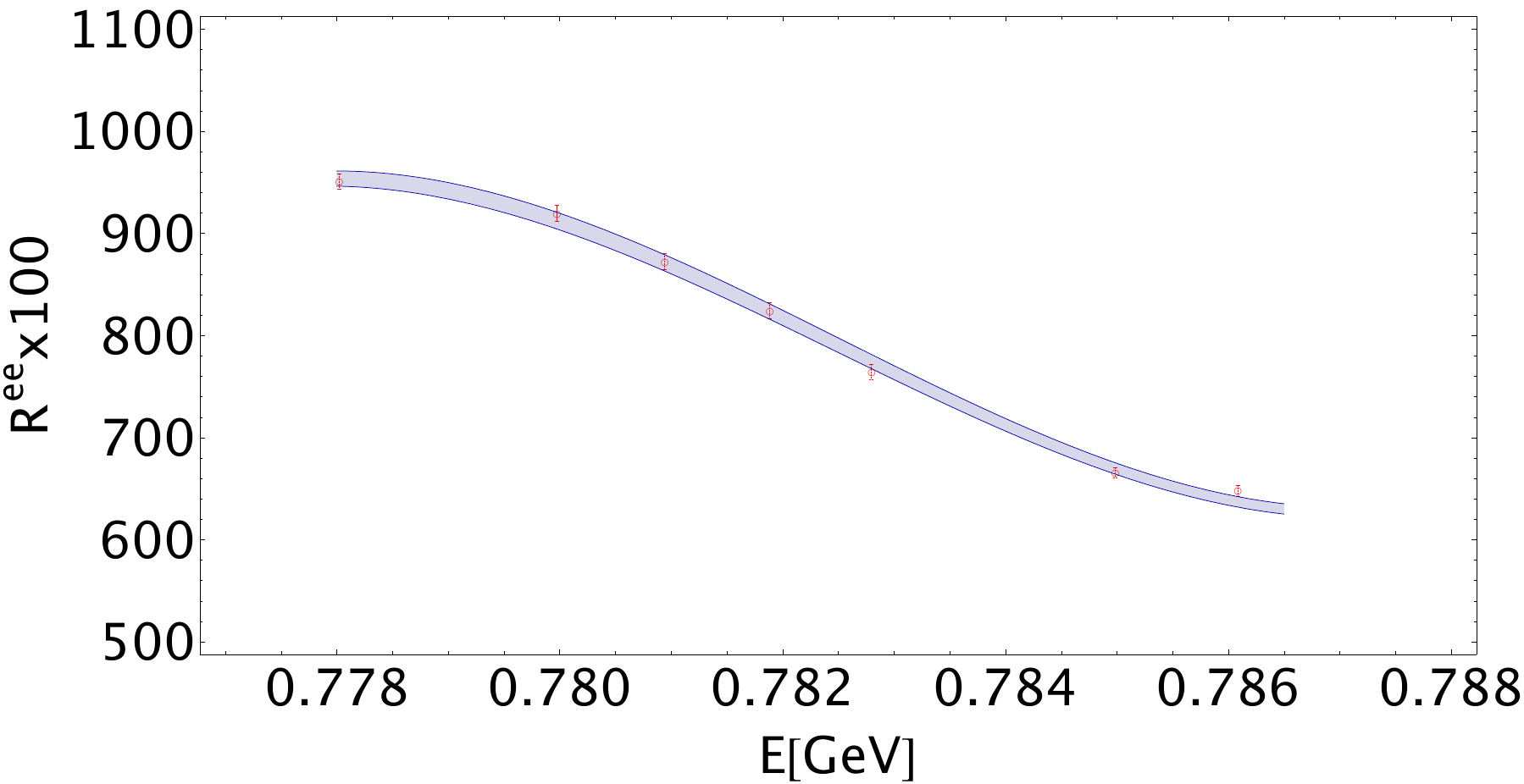}
\includegraphics[width=8cm]{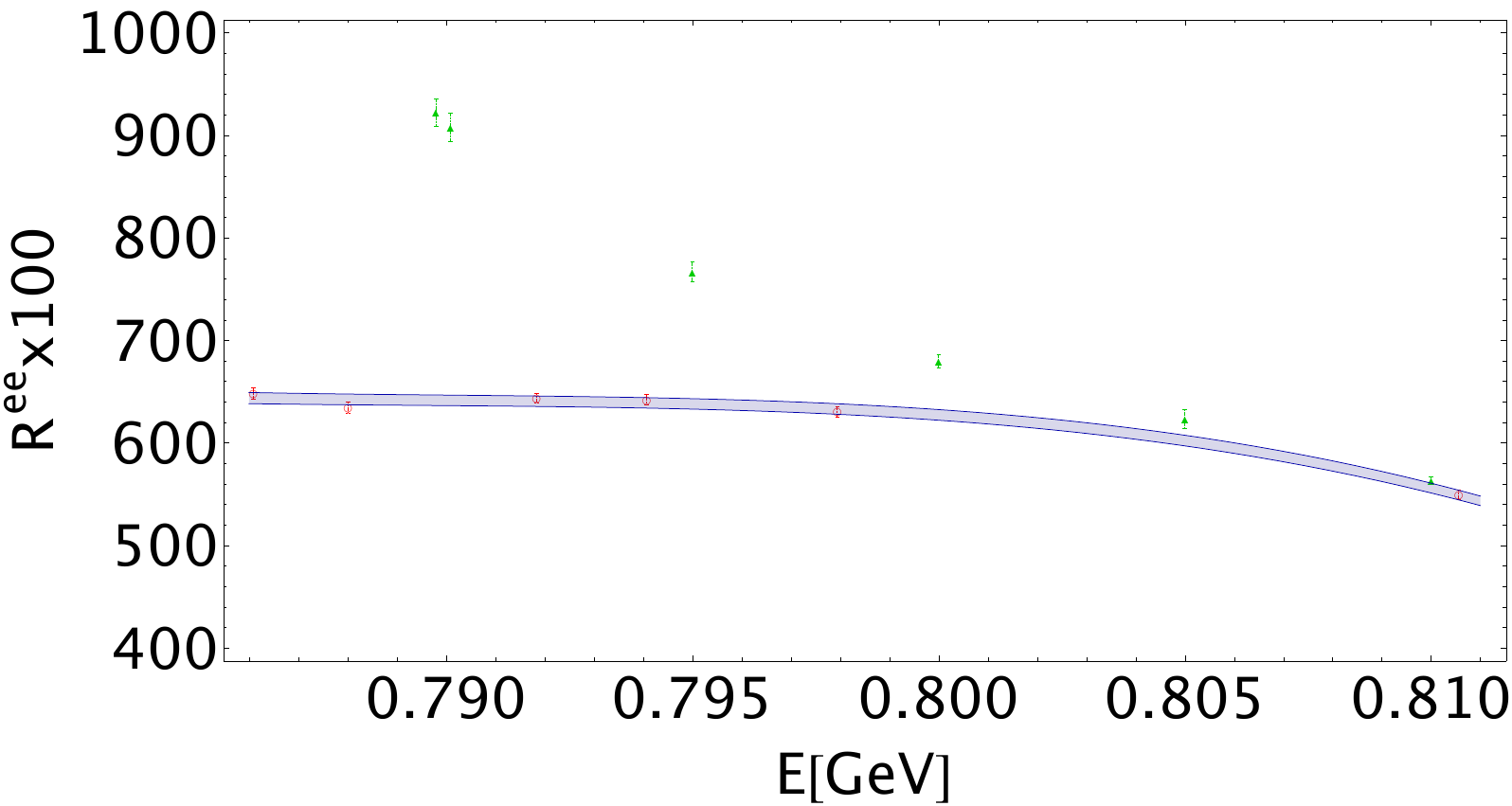}
\includegraphics[width=10cm]{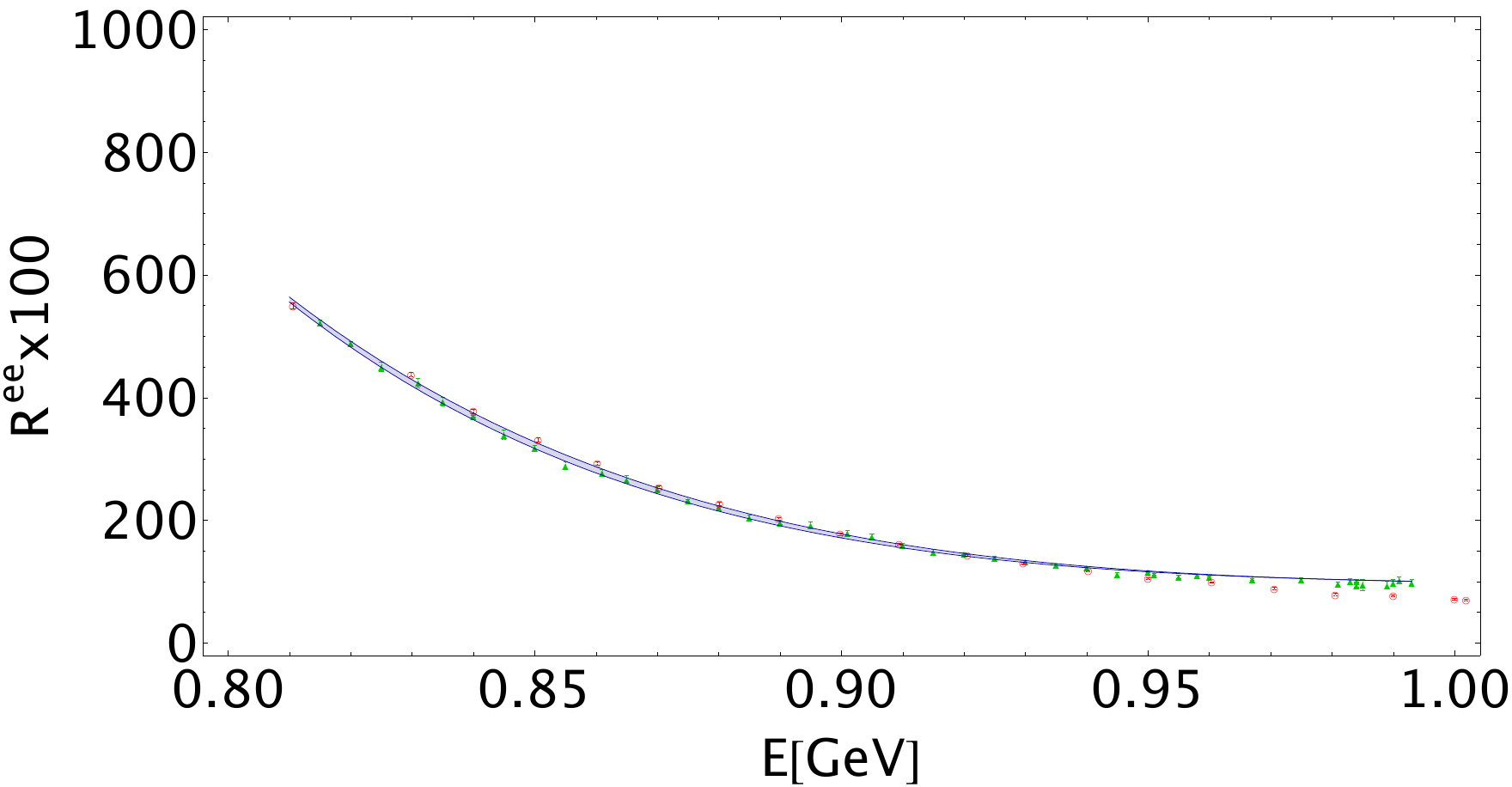}
\caption{\footnotesize  Fit of the data in the region 0.776 to 0.993 GeV. } \label{fig:rho3}
\end{center}
\vspace*{-0.5cm}
\end{figure}

Though our minimal BW parametrization is in agreement with previous determination given in Fig.\,\ref{fig:amu}, we can see in Fig.\,\ref{fig:rho} that there are regions which need to be improved. This is necessary for the high-precision determination of $a_\mu$. Notice that, within the modest accuracy of the results in previous sections, this improvement only affects slightly these results.  In so doing , we subdivide the region below 0.993 GeV into 6 subregions in units of GeV\,: 
\beq
[2m_\pi\to 0.5],\,\,\,[0.5 \to 0.6],\,\,\,[0.6\to 0.776],\,\,\,[0.776\to 0.786],\,\,\,[0.786 \to 0.810],\,\,\, [0.810\to .993].
\eeq
 We use the tail of the Breit-Wigner for [$2m_\pi\to$ 0.5] as shown in Fig.\,\ref{fig:rho} and different  polynomials within the FindFit Mathematica program with optimised $\chi^2$ to fit  the other regions from $0.5 \to .993$ GeV. We remind that the new CMD3 data\,\cite{CMD3}  are 2-3 times more precise than the one compiled by PDG[43] but the central values are approximately similar except in the low energy region where the CMD3 data are slightly higher.
 We show the analysis in Figs.\,\ref{fig:rho2} and \ref{fig:rho3}. Notice that, in the region 0.776 to 0.86 GeV, we have only considered the data from CMD3 (red). The effect of this region to $a_\mu$ from CMD3 data is $(206.5\pm 1.7)\times 10^{-11}$ while the one using the BW in Fig.\,\ref{fig:rho} gives $(205.7\pm 1.03)\times 10^{-11}$ such that the difference by using the two data is negligible. The different contributions are given in Table\,\ref{tab:amu1}. From the fit improvement the result in Eq.\,\ref{eq:compare} becomes :
 \beq
a_\mu^{\rho}\vert^{hvp}_{l.o}[0.6\to 0.88]=(377.4\pm 3.1)\times 10^{-10}.
\label{eq:compare2}
\eeq
 It is also informative to give our result  in the so-called intermediate region $[0.6\to 0.993]$ GeV:
 \beq
 a_\mu^{\rho}\vert^{hvp}_{l.o}[0.6\to 0.993]=(400.9\pm 3.2)\times 10^{-10}.
 \eeq
\subsection*{\hspace*{0.5cm}\d   $a_\mu^{I=1}\vert^{hvp}_{l.o}$ from .993 to 1.875 GeV}
The fits of these regions have been done in previous sections. The result is given in Table\,\ref{tab:amu1}.
\subsection*{\b $a_\mu$ from   $e^+e^-\to I=0$ hadrons below 1.875 GeV}
\subsection*{\hspace*{0.5cm}\d  $\omega$ and $\phi$ mesons contributions}
We estimate the $\omega$ and $\phi$ mesons contribution using a narrow width approximation (NWA). It reads\,\cite{SNamu}:
\beq
a_\mu^{I=0}\vert^{hvp}_{l.o}= \frac{3}{\pi}\sum_{R=\omega,\phi}\frac{\Gamma^{ee}_R}{M_R} K_\mu(M_R^2).
\eeq
Using\,\cite{PDG} :
\bea
\Gamma^{ee}_\omega &=& (64\pm 1.9\pm 0.96)\times 10^{-2}~{\rm keV},\,\,\,\,\,\,\,\,\,\,\,\Gamma^{ee}_\phi = (126.6\pm 1.4\pm 0.4)\times 10^{-2}~{\rm keV}\nnb\\
M_\omega&=&(782.66\pm 0.13)~{\rm MeV},\,\,\,\,\,\,\,\,\,\,\,   \,\,\,\,\,\,\,\,\,\,\,  \,\,\,\,\,\,\,\,\,\, M_\phi=(1019.461\pm 0.016)~{\rm MeV},
\eea
the result is given in Table\,\ref{tab:amu1}.
\subsection*{\hspace*{0.5cm}\d   $a_\mu^{I=0}\vert^{hvp}_{l.o}$ from .993 to 1.5 GeV}
We add the $I=0$ contribution  substracted  for the $I=1$ analysis in the previous sections by using  the $SU(3)$ relation. We obtain the value in Table\,\ref{tab:amu1}. 

\subsection*{\hspace*{0.5cm}\d   $a_\mu^{I=0}\vert^{hvp}_{l.o}$ from 1.5 to 1.875 GeV}
We consider the contributions of the $\omega(1650)$ and $\phi(1680)$. We use the parameters deuced from PDG22\,\cite{PDG}:
\bea
\omega(1650)&:&  M_\omega=(1670\pm 30)\,{\rm MeV},\,\,\,\,\,\,        \,\,\,\,\,  \,\,\,\Gamma^{ee}_\omega=1.35\pm 0.14\,{\rm keV},\,\,\,\,\,\,  \,\,\,\,\,\,  \,\,\Gamma^{h}_\omega=315\pm 35\,{\rm MeV} \nnb\\
\phi(1680)&:&  M_\phi=(1680\pm 20)\,{\rm MeV},\,\,\,\,\,\,   \,\,\,     \,\,\,\,\,  \,\,\,\Gamma^{ee}_\phi=0.18\pm 0.06\,{\rm keV},\,\,\,\,\,\,  \,\,\,\,\,\,  \,\,\Gamma^{h}_\psi=150\pm 50\,{\rm MeV}.
\eea
Using a Breit-Wigner (BW) parametrization, we  obtain the value in Table\,\ref{tab:amu1}. 

\subsection*{\b   $a_\mu\vert^{hvp}_{l.o}$ from 1.875 to 2 GeV}
\begin{figure}[hbt]
\begin{center}
\includegraphics[width=8.cm]{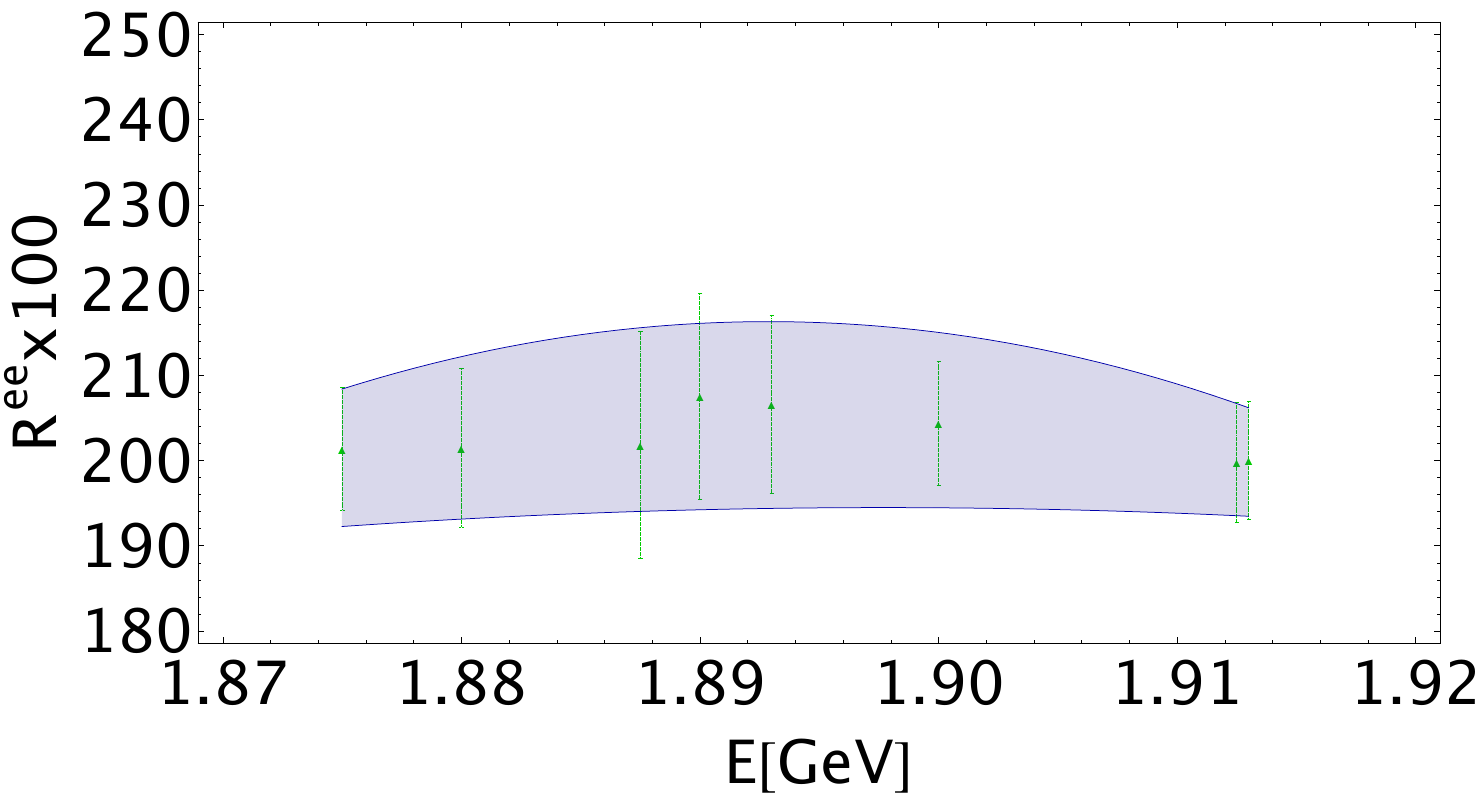}
\includegraphics[width=8.cm]{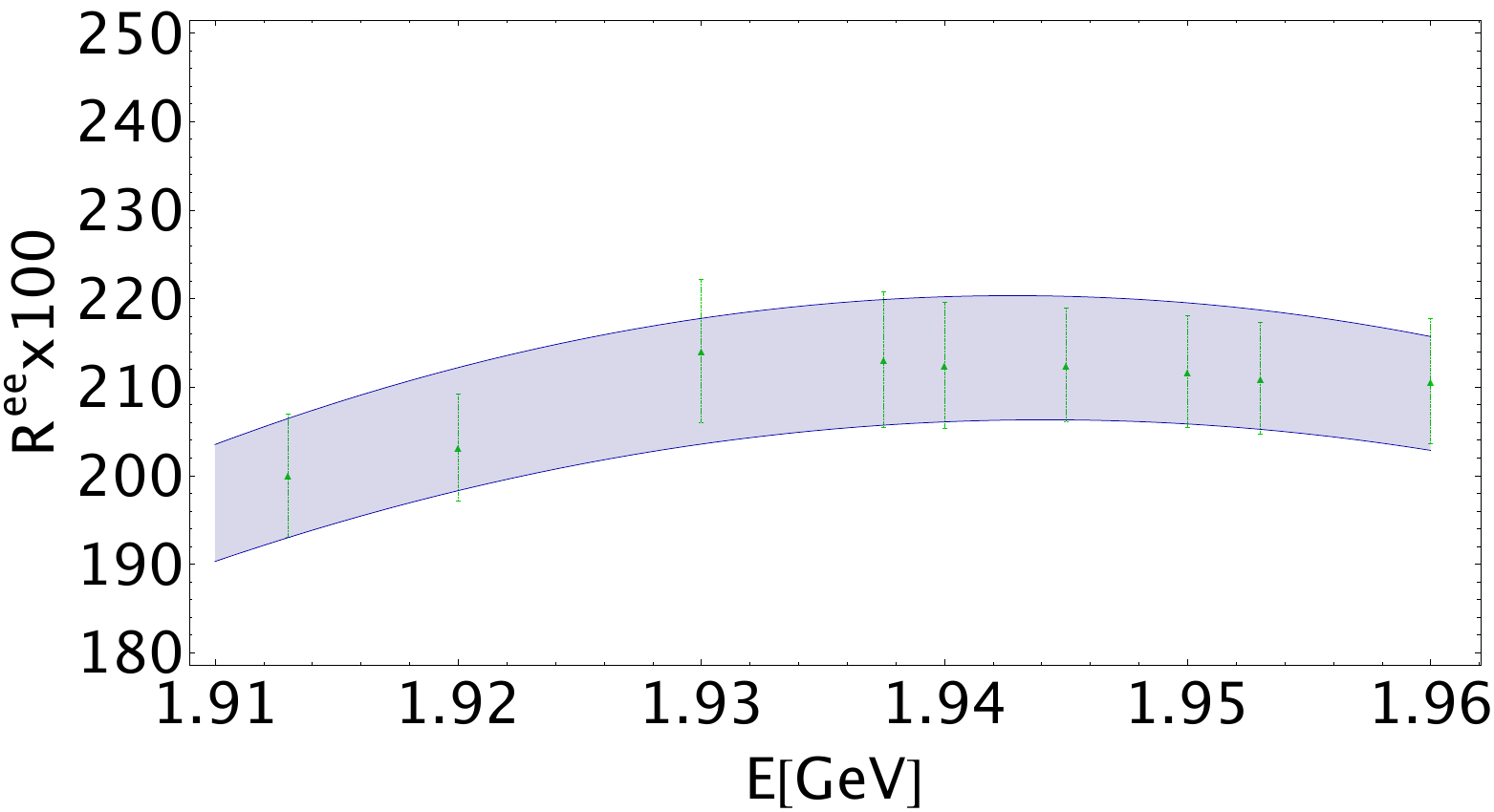}
\includegraphics[width=10cm]{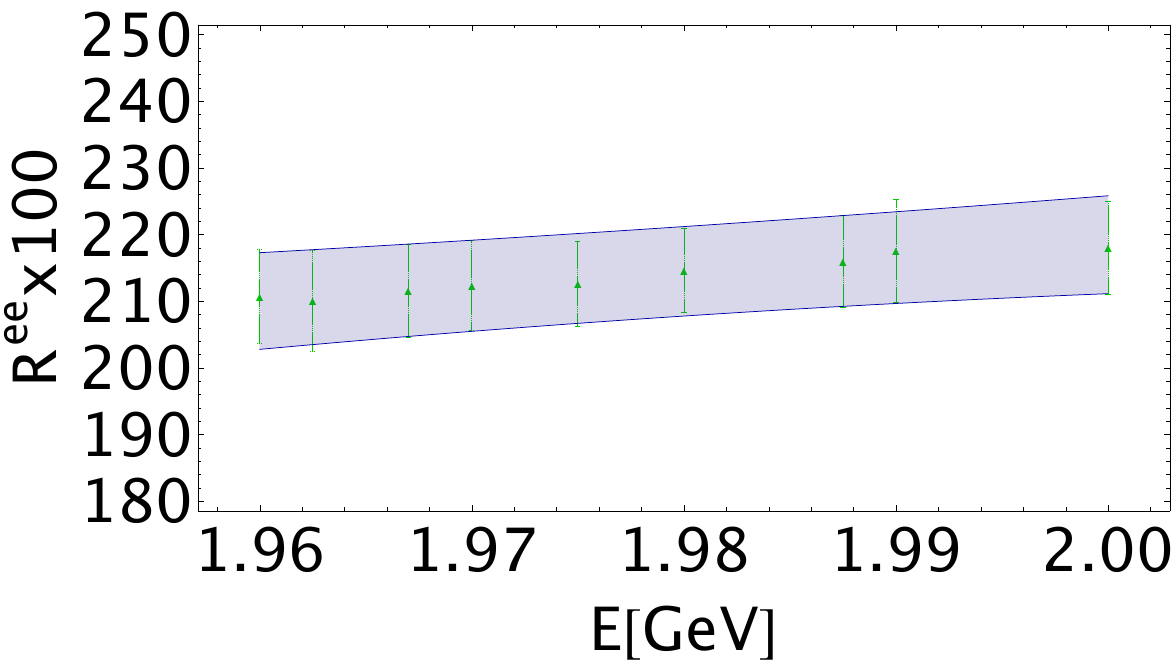}
\caption{\footnotesize  Fit of the data in the region 1.875 to 2 GeV. } \label{fig:fit-R5}
\end{center}
\vspace*{-0.5cm}
\end{figure}

To get this contribution, we divide the data into three regions and fit with polynomials using the optimized $\chi^2$ Mathematica program FindFit. The fits of the data are shown in Fig.\ref{fig:fit-R5}. 
\subsection*{\b   $a_\mu\vert^{hvp}_{l.o}$ from 2 to 3.68 GeV}
\begin{figure}[hbt]
\begin{center}
\includegraphics[width=13cm]{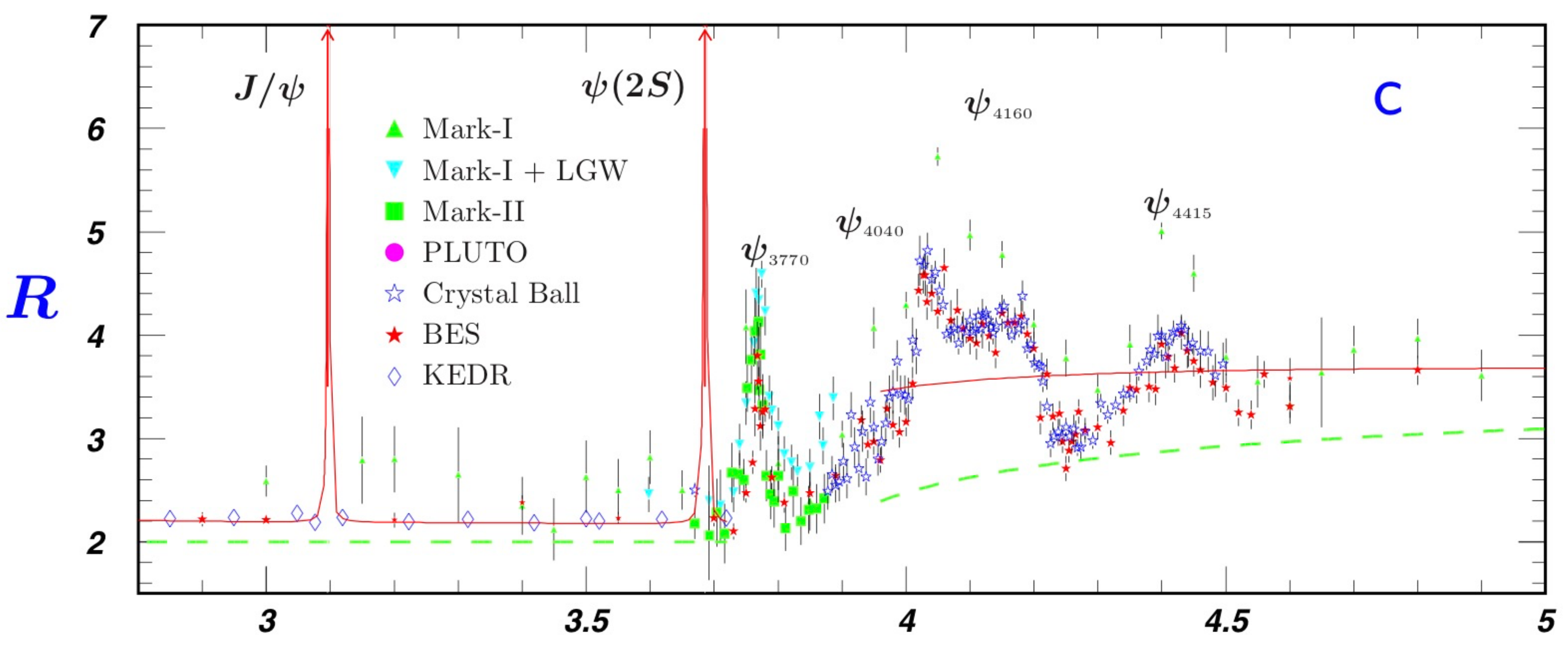}
\caption{\footnotesize  $e^+e^-\to$ Hadrons data in the charmonium region from 2 to 5 GeV from PDG\,\cite{PDG}. } \label{fig:charm}
\end{center}
\vspace*{-0.5cm}
\end{figure}

The data in this region are well fitted by the QCD expression of the spectral function for 3 flavours as one can see from Fig.\,\ref{fig:charm} given by PDG. 
To the massless PT expression known to order $\alpha_s^4$, we  include the quark and gluon condensates of dimensions $D=4,6$  (see  Eqs.\,\ref{eq:pt} to \ref{eq:d6}).  We add the quadratic  $m^2_s$-corrections to order $\alpha_s^3$ and the quartic $m_s^4$-mass corrections to order $\alpha_s^2$\,\cite{CHETmass}.  These quark mass corrections read in the $\overline{MS}$-scheme:
\bea
d_{2m}&=& -\frac{\overline{m_s}^2}{t}\ga 12\,a_s+104.833\,a_s^2+541.735\,a_s^3\dr,\nnb\\
d_{4m}&=&\frac{\overline{m_s}^4}{t^2}\Bigg{[} -6(1+\frac{11}{3}a_s) +a_s^2\Big{[}\, n_f\ga\,\frac{\rm L_m}{3}-1.841\dr -\frac{11}{2}\,{\rm L_m}+136.693+12-0.475-{\rm L_m}\Big{]}\Bigg{]},
\label{eq:m4}
\eea
where $\overline{m}$ is the running quark mass and ${\rm L_m}\equiv {\rm Log}({\overline m_s^2/t})$. In the numerical analysis, we shall also introduce the RGI quark mass defined as\,\cite{FNR}\,:
\beq
\overline{m_s}(\nu)=\hat m_s\,(-\beta_1 a_s(\nu))^{-\gamma_1\beta_1}\Big{[} 1+{\cal O}(a_s)\Big{]},
\eeq
where $\gamma_1=2$ is the quark mass anomalous dimension and $\beta_1=-(11-2n_f/3)/2$ the first coefficient of the $\beta$-function. The expression to order $a_s^2$ which we shall use can be found e.g. in Ref.\,\cite{SNB2}. We shall use $\hat m_s=$ 116(6) MeV\,\cite{SNREV1}. 

\subsection*{\b   $a_\mu\vert^{hvp}_{l.o}$ from charmonium}
\subsection*{\hspace*{0.5cm} \d The $J/\psi(1S),\psi(2S)$ and $\psi(3773)$ contributions}
We estimate their contributions using a NWA and the values of the masses and widths from PDG\,\cite{PDG}:
\bea
\psi(1S)&:&  M_\psi=(3069.9\pm 0.006)\,{\rm MeV},\,\,\,\,\,\,        \,\,\,\,\,  \,\,\,\Gamma^{ee}_\psi=(5.53\pm 0.10)\,{\rm keV},\,\,\,\,\,\,  \,\,\,\,\,\,  \,\,\Gamma^{h}_\psi=(92.6\pm 1.0)\,{\rm keV}. \nnb\\
\psi(2S)&:&  M_\psi=(3681.1\pm 0.06)\,{\rm MeV},\,\,\,\,\,\,   \,\,\,     \,\,\,\,\,  \,\,\,\Gamma^{ee}_\psi=(2.33\pm 0.08)\,{\rm keV},\,\,\,\,\,\,  \,\,\,\,\,\,  \,\,\Gamma^{h}_\psi=(294\pm 8)\,{\rm keV}. \nnb\\
\psi(3773)&:&  M_\psi=(3773.7\pm 0.4)\,{\rm MeV},\,\,\,\,   \,\,\, \,\,\,   \,\,\,\,\,  \,\,\,\Gamma^{ee}_\psi=(0.26\pm 0.02)\,{\rm keV}, \,\,\,\,\,\,  \,\,\,\,\,\,  \,\,\Gamma^{h}_\psi=(27.2\pm 1.0)\,{\rm MeV}.
\eea
The result is given in Table\,\ref{tab:amu1}.
\begin{figure}[hbt]
\begin{center}
\includegraphics[width=7.cm]{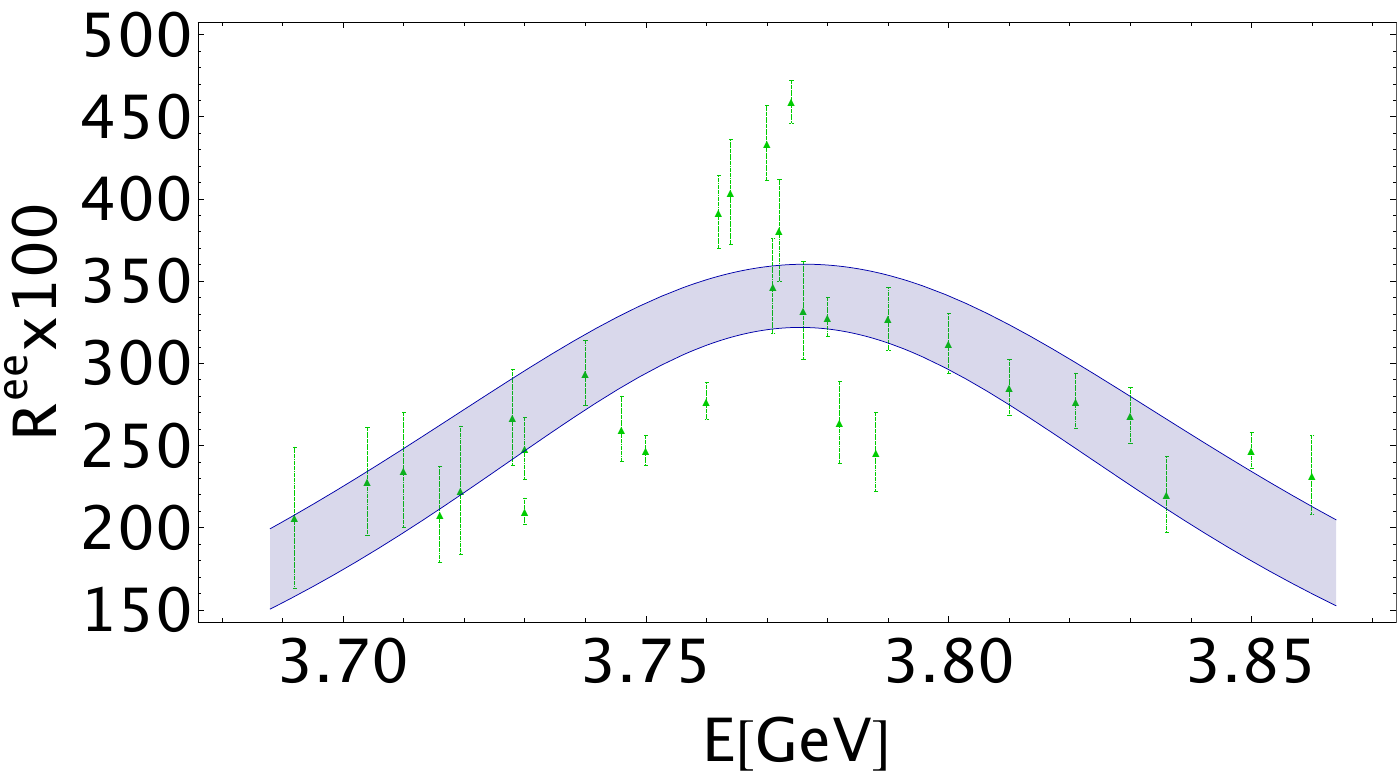}
\includegraphics[width=7.cm]{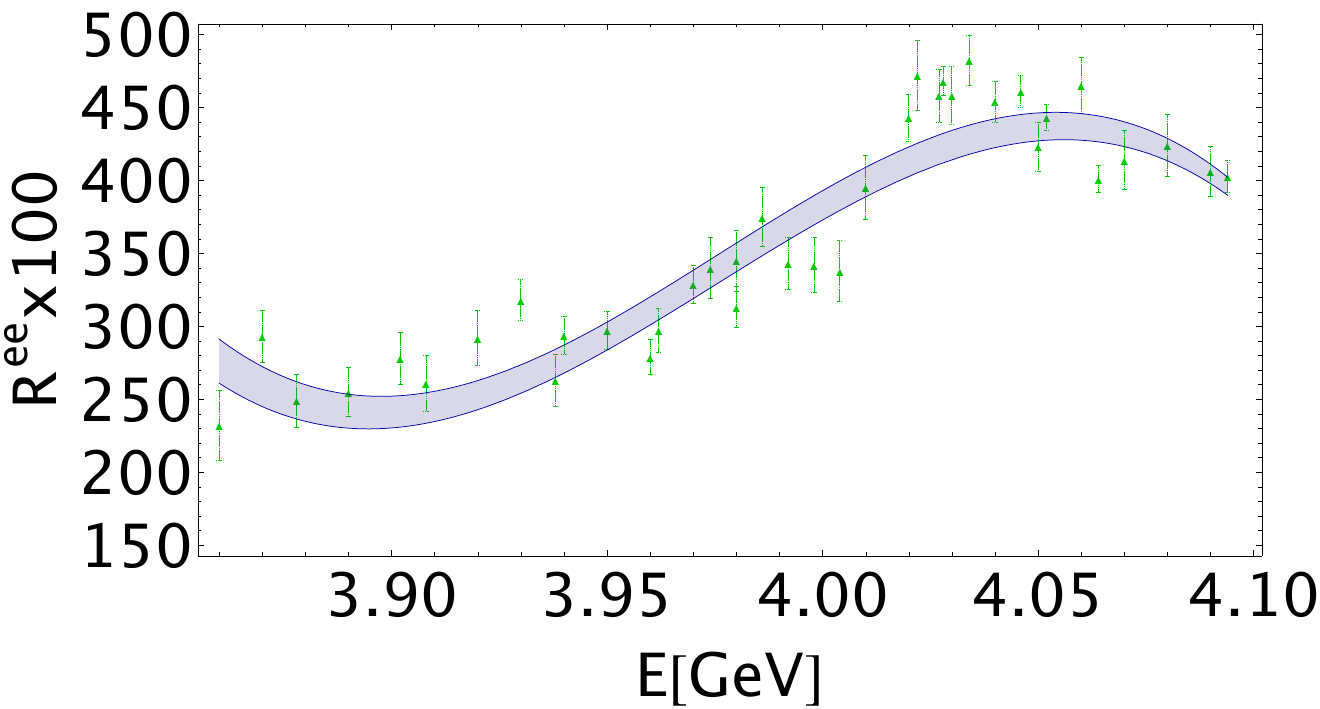}\\
\includegraphics[width=7.cm]{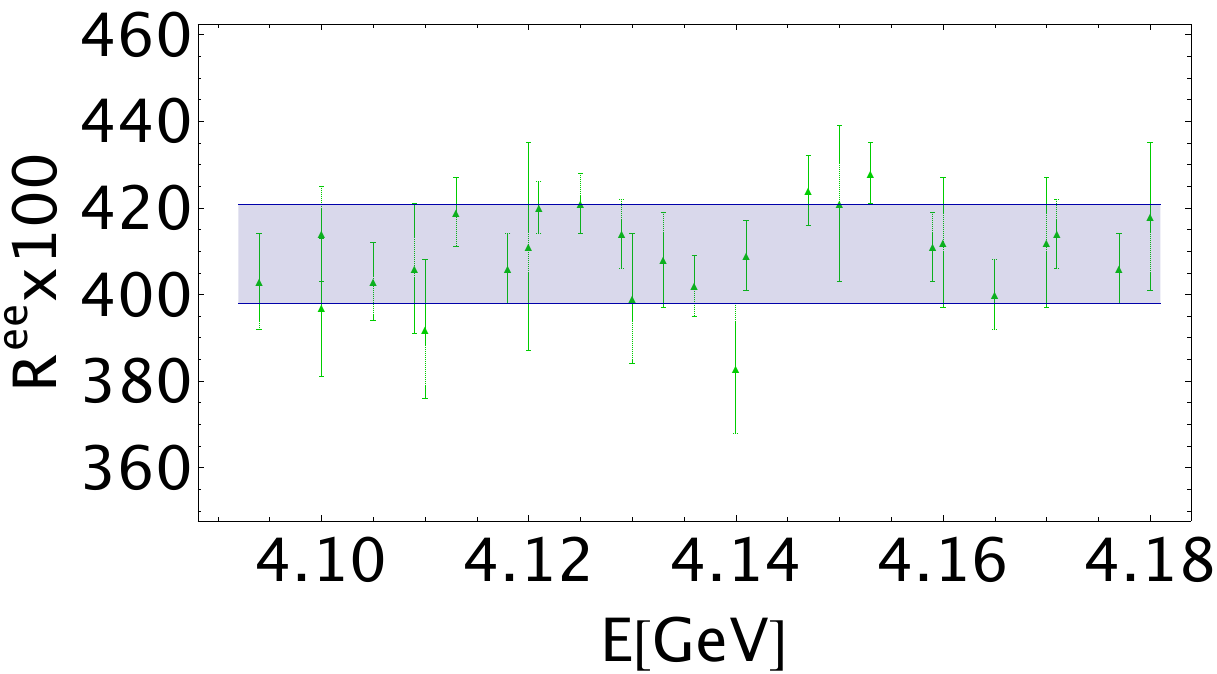}
\includegraphics[width=7.cm]{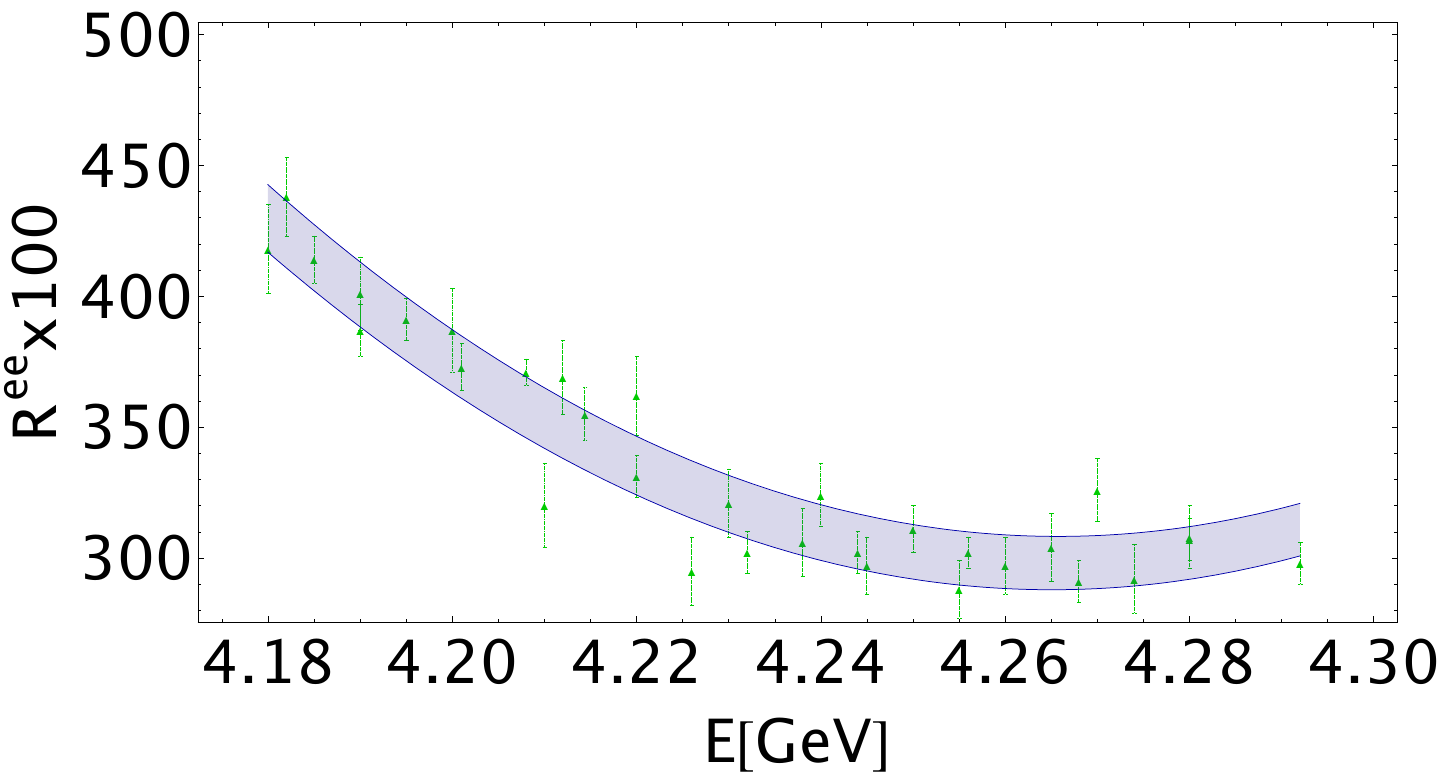}\\
\includegraphics[width=8cm]{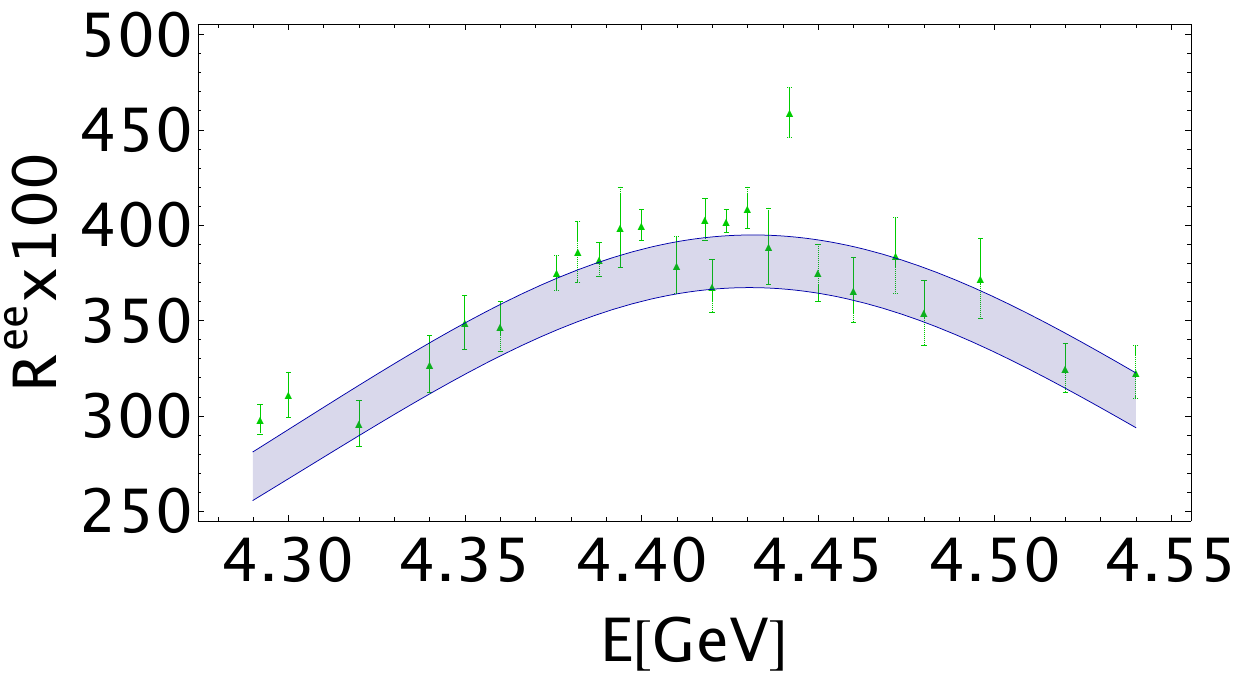}
\caption{\footnotesize  Fits of the $e^+e^-\to$ Hadrons data in the charmonium region from 3.68 to 4.55 GeV from PDG\,\cite{PDG}. } \label{fig:fit-R7}
\end{center}
\vspace*{-0.5cm}
\end{figure} 
\subsection*{\d Region from 3.68 to 4.55 GeV  }
We divide this region into 5 subregions as shown in Fig.\,\ref{fig:fit-R7}. The region between 3.68 to 3.86 is better fitted using a Breit-Wigner while in the others,  we use polynomials. The different contributions are shown in Table\,\ref{tab:amu1}. 
\subsection*{\vspace*{0.5cm}\d Region from 4.55 to 10.50 GeV }
\begin{figure}[hbt]
\begin{center}
\includegraphics[width=13cm]{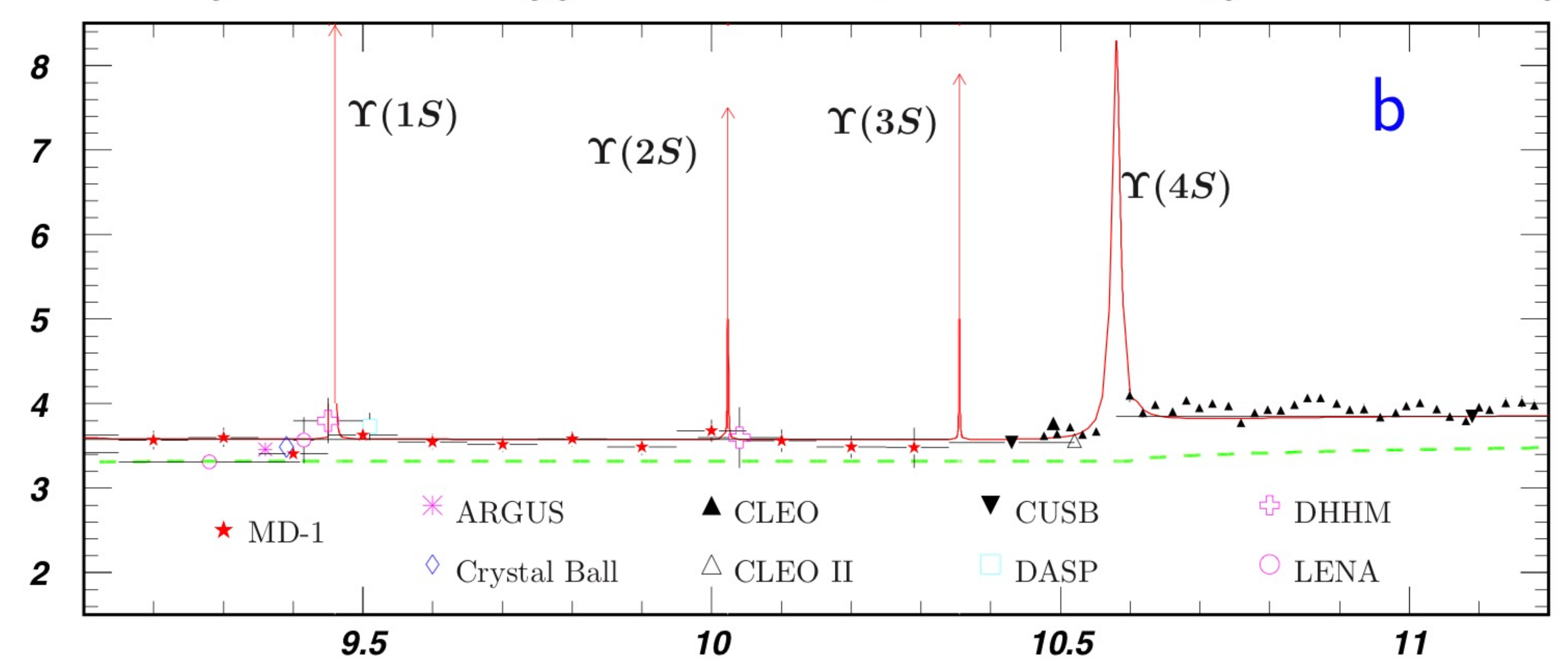}
\caption{\footnotesize  $e^+e^-\to$ Hadrons data in the bottomium region from 9 to 10.5 GeV from PDG\,\cite{PDG}. } \label{fig:bottom}
\end{center}
\vspace*{-0.5cm}
\end{figure}

As shown in Figs.\,\ref{fig:charm} and \ref{fig:bottom}, the data without resonance peaks are well fitted by QCD for 4 flavours. We add to the previous QCD expressions the charm contributions with $m_c^2$ and $m_c^4$  mass corrections where $\overline{m}_c(\overline{m}_c)=$ 1266(6) MeV\,\cite{SNREV1}. The contribution is given in Table\,\ref{tab:amu1}. 
\subsection*{\b   $a_\mu\vert^{hvp}_{l.o}$ from bottomium}
\subsection*{ \vspace*{0.5cm} \d $\Upsilon(1S\to 11.02)$ contributions}
We use a NWA to estimate their contributions with:
\bea
\Upsilon(1S)&:&  M_\Upsilon=(9460.30\pm 0.26)\,{\rm MeV},\,\,\,\,\,\,        \,\,\,\,  \,\,\,\Gamma^{ee}_\Upsilon=(1.286\pm 0.067)\,{\rm keV},\,\,\,\,\,\,  \,\,\,\,\,  \,\,\Gamma^{h}_\Upsilon=(954.02\pm 1.25)\,{\rm keV} ,\nnb\\
\Upsilon(2S)&:&  M_\Upsilon=(10023.26\pm 0.31)\,{\rm MeV},\,\,\,\,\,\,        \,\,\,\,  \,\,\,\Gamma^{ee}_\Upsilon=(0.611\pm 0.071)\,{\rm keV},\,\,\,\,\,\,  \,\,\,\,\,  \,\,\Gamma^{h}_\Upsilon=(31.98\pm 2.63)\,{\rm keV} ,\nnb\\
\Upsilon(3S)&:&  M_\Upsilon=(10355.2\pm 0.5)\,{\rm MeV},\,\,\,\,\,\,        \,\,\,\,  \,\,\,\Gamma^{ee}_\Upsilon=(0.443\pm 0.057)\,{\rm keV},\,\,\,\,\,\,  \,\,\,\,\,  \,\,\Gamma^{h}_\Upsilon=(20.32\pm 1.85)\,{\rm keV} ,\nnb\\
\Upsilon(4S)&:&  M_\Upsilon=(10579.4\pm 1.2)\,{\rm MeV},\,\,\,\,\,\,        \,\,\,\,  \,\,\,\Gamma^{ee}_\Upsilon=(0.322\pm 0.042)\,{\rm keV},\,\,\,\,\,\,  \,\,\,\,\,  \,\,\Gamma^{h}_\Upsilon=(20.5\pm 2.5)\,{\rm keV} .
\eea
\subsection*{ \vspace*{0.5cm} \d QCD continuum contribution from  10.59 GeV to $2m_t$}
We add the $b$-quark contribution to the previous QCD expression where $b$-quark mass corrections to order $a_s^3\overline{m}_b^2/t$ and $a_s^2\overline{m}_b^4/t^2$ are included. We use\,\: $\overline{m}_b(\overline{m}_b)=$ 4202(8) MeV\,\cite{SNREV1}.
For the analysis, we consider the region from 10.59 MeV to $2m_t$ just after the $\Upsilon(4S)$ where the QCD continuum is expected to smear the  $\Upsilon(10860, 11020)$  and some eventual  higher resonances. The result is given in Table\,\ref{tab:amu1}. 
\subsection*{ \vspace*{0.5cm} \d QCD continuum contribution from  $2m_t\to\infty$}
Due to the heaviness of the top quark mass, we shall use the approximate Schwinger formula near the $\bar tt$ threshold
for a much better description of the spectral function up to order $\alpha_s$ due to the top quark:
\beq
R_t^{ee}=\frac{4}{3} v\frac{(3-v^2)}{2}\Big{[} 1+\frac{4}{3}\alpha_sf(v)\Big{]},
\eeq
with:
\beq
f(v)=\frac{\pi}{2\, v}-\frac{(3+v)}{4}\ga \frac{\pi}{2}-\frac{3}{4\pi}\dr\,:\,\,\,\,\,\, v=\ga 1-\frac{m_t^2}{t}\dr^{1/2}.
\eeq
Here $m_t$ is the on-shell top quark mass which we fix to be\,\cite{PDG}\,\footnote{We should note that the definition of the top quark mass from different experiments is still ambiguous.}: 
\beq
\overline{m}_t(\overline{m}_t^2)=(172.7\pm 0.3)\,{\rm  GeV},
\label{eq:mt}
\eeq
 from some  direct measurements.
We add to this expression the  one due to $\alpha_s^2$ and $\alpha_s^3$ given in Eq.\,\ref{eq:m4} within the $\overline {MS}$-scheme.  For this $\overline {MS}$-scheme expression  in terms of running mass, we need the value of the RGI top mass:
\beq
\hat m_t=(254\pm 0.4)\,{\rm GeV},
\eeq
deduced from Eq.\,\ref{eq:mt}.
Adding the above expressions to the ones in the previous sections, we obtain the result given in Table\,\ref{tab:amu1}. 
\section{Conclusion for $a_\mu\vert^{hvp}_{l.o}$ and extension to $a_\tau\vert^{hvp}_{l.o}$}
\subsection*{\b Results for $a_\mu\vert^{hvp}_{l.o}$} 
We have used  the sum  exclusive $e^+e^-\to$ hadrons data compiled by PDG $\oplus$ some resonances to extract the 
lowest order vacuum polarization contribution to the muon anomaly. We find from Table\,\ref{tab:amu1}\,:
\beq
a_\mu\vert^{hvp}_{l.o}=(7036.5\pm 38.9)\times 10^{-11},
\eeq
where the largest contribution and error come (as expected) from the $\rho$-meson low-energy one. It is amazing that the central value remains stable when comparing it with our old results in Refs.\,\cite{CALMET,SN76,SN78,SNamu} though the accuracy has increased by a huge factor of about 24 (!) thanks to the experimental efforts for improving the data during about half century !  The total sum is slightly larger  than the recent data based determinations in Refs.\,\cite{NOMURA,DAVIER2}\,\footnote{Some earlier references are quoted in these papers and in\,\cite{SNamu}.} but in better agreement with a recent analysis of $\tau$-decay data\,\cite{MIRANDA}  and with some recent lattice results\,\cite{RAF,GM2,LATTICE}.  Using this value into Table 8 of Ref.\,\cite{KNECHT} and in Table 1 of Ref.\,\cite{GM2}where some other sources of contributions are reviewed, we deduce:
\beq
a_\mu^{exp}= 1165911898(42)\times 10^{-11}.
\eeq
This leads to:
\beq
\Delta a_\mu\equiv a_\mu^{exp}-a_\mu^{th} = (142\pm 42_{th}\pm 41_{exp})\times 10^{-11},
\label{eq:amu-sm}
\eeq
which reduces the tension between experiment and the SM prediction. We have used the experimental average from BNL\,\cite{BNL} and FNAL\,\cite{FNAL}\,:
\beq
a_\mu^{exp}= 116592040(41)\times 10^{-11}.
\eeq
\subsection*{\b Some comments on the determination of  $a_\mu\vert^{hvp}_{l.o}$} 
\d  Instead of our simple Breit-Wigner parametrization of pion form factor done in previous sections, we have subdivided the region below 0.993 GeV to have a better fit of the data which is necessary for the extraction of the lepton anomalies. 
We have compared our result in this region with  the ones from different experimental groups quoted in Fig\,\ref{fig:amu}.

\d Our analysis differs from the most recent ones in Refs.\,\cite{NOMURA,DAVIER2} as we fit the sum of exclusive modes except the $\rho$-meson. For the narrow resonances we use the NWA.

\d We notice that compared to the existing analysis, we have also studied in details the contributions from the heavy quark sectors taking into account all possible resonances and, in particular, analyzed carefully the charmonium region. 

\d We have also carefully parametrized the QCD continuum contributions taking into account higher order PT quark mass corrections and the non-perturbative ones. 

\d Comparing our results with the most recent ones in Refs.\,\cite{DAVIER2,NOMURA}, we found that in the low-energy region below 1.875 GeV, our result is higher by about $(100-136)\times 10^{-11}$ than the ones in these references. This can be due to the pion form factor where we use the new data of CMD3 $\oplus$ PDG which leads to higher value of $a_\mu$ than some other determinations as shown in Fig.\ref{fig:amu} for e.g. the region $0.60\leq \sqrt{t}\leq 0.88$ GeV.

\d In the high-energy region $\sqrt{t}\geq 1.875$ GeV, our result is $(545.3\pm 2.2)\times 10^{-11}$ which is  about the same as  the one $(535.5\pm 7.0)\times 10^{-11}$ of Ref.\,\cite{NOMURA}  but smaller by about 38$\times 10^{-11}$ than $(583.6\pm 3.3)\times 10^{-11}$ in Ref.\,\cite{DAVIER2}. This difference is mainly due to  the choice of the QCD continuum threshold $\sqrt{t_c}= 1.8$ GeV in Ref.\,\cite{DAVIER2}  which is lower than the one found from the asymptotic coincidence of the two sides of ${\cal L}_0$ Laplace moment  in Eq.\,\ref{eq:tc} and requested by duality from FESR   in Eq. \ref{eq:fesr}. 

\subsection*{\b Extension to the determination of  $a_\tau\vert^{hvp}_{l.o}$} 
\d The muon analysis has been extended to the case of the $\tau$-lepton by simply changing the lepton mass. The total sum is from Table\,\ref{tab:amu1}\,:
\beq
a_\tau\vert^{hvp}_{l.o}=(3494.8\pm 24.7)\times 10^{-9},
\eeq
which we consider as an improvement of the {\it pioneer determination} of this quantity and of $a_\tau$ in Ref.\,\cite{SN78}. 

\d One can notice that the relative weight of the light quarks over the heavy ones moves from 23 to 6 from $\mu$ to $\tau$ indicating that a measurement of the $\tau$-anomaly will probe higher energy region thanks to the behaviour of the QED kernel function  given in Eq.\,\ref{eq:kmu}. 

\d Our result is slightly lower than the one of Ref.\,\cite{NOMURA} where the origin may come from the $I=0$ light mesons region. 

\section{Determination of $ \Delta \alpha_{had}^{(5)} (M_Z^2)$}
We conclude the paper by updating our previous determination of $\Delta \alpha_{had}^{(5)} (M_Z^2)$ in Ref.\,\cite{SNalfa}. The hadronic contribution to this quantity can be expressed as:
\beq
\Delta \alpha_{had}^{(5)} (M_Z^2)=-\ga \frac{\alpha}{3\pi}\dr M^2_Z\int_{4m_\pi^2}^\infty \frac{R_{ee}(t)}{t(t-M_Z^2)},
\eeq
where $R_{ee}$ is the ratio of the $e^+e^-\to$ Hadrons over the  $e^+e^-\to \mu^+\mu^-$ total cross-sections. 
The results from different regions are shown in Table\,\ref{tab:amu1} where at the  $Z_0$ pole, we take the principal value of the integral which we take as:
\beq
\int_{4m_\pi^2}^{(M_Z- \Gamma_Z/2)^2}\hspace*{-1cm} dt\,f(t)+\int_{(M_Z+ \Gamma_Z/2)^2}^{\infty}\hspace*{-1cm}dt\,f(t),
\eeq
where $\Gamma =2.5$ GeV is the total hadronic $Z$-width. We add to the QCD continuum contribution the one of the $Z$-pole estimated to be\,\cite{YND17}:
\beq
\Delta \alpha_{had}^{(5)}\vert_{M_Z}= 29.2\times 10^{-5}.
\eeq
Then, we obtain the total sum:
\beq
\Delta \alpha_{had}^{(5)} (M_Z^2)= (2766.3\pm 4.5)\times 10^{-5}. 
\eeq 
This value is comparable with the ones in the literature reviewed e.g. in Ref.\,\cite{JEGER}.  It improves and confirms our previous determination in Ref.\,\cite{SNalfa}. 
\section{Summary}
\subsection*{\b QCD parameters}

In the first sections, we have used the $I=1$ isovector component of the $e^+e^-\to$ Hadrons data to extract, from the ratio ${\cal R}_{10}$ of Laplace sum rules, the values of the four-quark and dimension $D=8$ condensates appearing in the OPE using as input the value of the gluon condensate $\la\alpha_s G^2\ra$ (see Eq.\,\ref{eq:res-d68}). 

\d  We found the value in Eq.\,\ref{eq:res-4q} where the one of the four-quark condensate is larger by a factor about 6 than the vacuum saturation estimate. We also show that a simultaneous use of the standard SVZ value of the gluon condensate and the vacuum saturation estimate of the four-quark condensate is inconsistent. 

\d Using the value of the $d=6$ and $d=8$ condensates into the ratio ${\cal R}_{10}$, we extract the value of $\la\alpha_s G^2\ra$ in Eq.\,\ref{eq:ag2} which is perfectly consistent with the one in Eq.\,\ref{eq:asg2} from heavy quarkonia.

\d Using the previous values of the condensates, we extract $\alpha_s$ from the lowest $\tau$-decay moment at different scale and found the values in Eq.\,\ref{eq:alphas} within FO/CI perturbation theory. This value is in good agreement with the ones from higher $\tau$-like decay moments within FO/CI\,\cite{PICH2}. 

\subsection*{\b SM-High-Precision parameters}
We complete our analysis by extracting some high-precision standard model (SM) parameters using the sum of exclusive $e^+e^-\to$ Hadrons data.

\d We obtain the values of the lowest order hadronic vacuum polarisation to the muon and tau anomalies $a_\mu\vert^{hvp}_{l.o}$ and $a_\tau\vert^{hvp}_{l.o}$ in Table\,\ref{tab:amu1}. 
\d We determine the hadronic contribution  to the running QED coupling $ \Delta \alpha_{had}^{(5)} (M_Z^2)$ in Table\,\ref{tab:amu1}. We discuss the sources of difference of our results with the most recent determinations.

Combined with some other contributions, these results are useful for a high-precision determination of the lepton anomalies $a_{\mu,\tau}$ and the running QED coupling $\alpha(M_Z^2)$.  In particular, using  the other sources of contributions to $a_\mu^{th}$ from Table 8 of Ref.\,\cite{KNECHT} and the updated review in Ref.\cite{GM2}, we obtain the result in Eq.\,\ref{eq:amu-sm} which decreases the tension between the SM prediction and experiment.  

\section{Notes added}
\subsubsection*{\b Muon anomalous magentic moment $a_\mu$}
After the publication of this work, a new experimental measurement of the  anomaly: $a_\mu\equiv \frac{1}{2}(g-2)_\mu$ of a positive charged muon has  been published\,\cite{MG2}. This has lead to a new value of the comparison between the SM prediction and experiment\,\cite{SN23}\,:
\beq
\Delta a_\mu\equiv a_\mu^{exp}-a_\mu^{th} = (143\pm 42_{th}\pm 22_{exp})\times 10^{-11}, 
\eeq
which indicates about 3\,$\sigma$ discrepancy. This number updates the one quoted in the Abstract and in Eq. 82. 

\subsubsection*{\b QCD power corrections and $\alpha_s$}
The estimate of power corrections has been recently improved by a combined use of the ratio of Laplace Sum Rule (LSR) and $\tau$-like decay high-moments\,\cite{SN24}. This analysis leads to the values of the gluon $\la\alpha_s G^2\ra$,  and $d_6$, $d_8$ condensates given in Table 1 of Ref.\,\cite{SN24}.  With the new values of $d_6$ and $d_8$,  the ones of  $\alpha_s(M_\tau)$ to order $\alpha_s^4$ quoted in the abstract and in Eqs. 56 to 60 becomes the one in Eqs. 30 and 31 of Ref.\,\cite{SN24}. The increase of the value of $d_6$ has slitghly decreased the one of $\alpha_s$. To order $\alpha_s^4$, one obtains:
\bea
\alpha_s(M_\tau) &=& 0.3081(49)_{fit}(71)_{\alpha_s^5} ~~~\lrar~~~\alpha_s(M_Z) = 0.1170(6)(3)_{evol} ~~~~~~   {\rm (FO)}\nnb\\
&=& 0. 3260(47)_{fit}(62)_{\alpha_s^5} ~~~\lrar~~~\alpha_s(M_Z) = 0.1192(6)(3)_{evol} ~~~~~~{\rm (CI)},
\label{eq:as-foci}
\eea
where the sum of non-perturbative corrections at the $\tau$-mass  is: $\delta^{(NP)}=(3.8\pm 0.8)\times 10^{-2} $ (Table 3 of Ref.\,\cite{SN24}). The previous values lead to the mean:
\beq
\alpha_s(M_\tau)\vert_{SVZ} = 0.3194(52)(113)_{syst} ~~~\lrar~~~\alpha_s(M_Z)\vert_{SVZ} = 0.1184(15)(3)_{evol},
\label{eq:as-final}
\eeq
where the systematics is the distance between the mean central value to the FO/CI ones. 
\section*{Acknowledgement}
We thank 
Antonio Pich and  Eduardo de Rafael for some correspondences.
\newpage

\end{document}